\documentclass[a4paper,fleqn]{cas-dc}

\usepackage[authoryear]{natbib}

\usepackage{amssymb}
\usepackage{lipsum}
\usepackage{caption}
\usepackage{subcaption}
\usepackage{amsmath}
\usepackage{cleveref}
\crefname{section}{Section}{Sections}
\crefname{figure}{Fig.}{Figs.}
\crefname{equation}{Eq.}{Eqs}
\crefname{table}{Table}{Tables}
\usepackage{tabularx}
\usepackage{textgreek}
\usepackage{gensymb}
\usepackage{graphicx}
\usepackage[export]{adjustbox}
\usepackage{placeins}
\usepackage{xurl} % Allows breaking URLs at any point
\def\tsc#1{\csdef{#1}{\textsc{\lowercase{#1}}\xspace}}
\tsc{WGM}
\tsc{QE}
\begin{document}
\let\WriteBookmarks\relax
\def\floatpagepagefraction{1}
\def\textpagefraction{.001}
\shorttitle{}
\shortauthors{Wei QIN et~al.}
%\begin{frontmatter}

\title [mode = title]{Three-dimensional numerical study on hydrogen bubble growth at electrode}                      
\tnotemark[1]

\tnotetext[1]{This project has received funding from the European Research Council (ERC) under the
European Union’s Horizon 2020 research and innovation programme (grant agreement number 883849).}

%\tnotetext[2]{The second title footnote which is a longer text matter
   %to fill through the whole text width and overflow into
   %another line in the footnotes area of the first page.}

\author[1]{Wei QIN}[orcid=0009-0004-1646-478X]
\cormark[1]
%\fnmark[1]
\ead{wei.qin@sorbonne-universite.fr}
%\ead[url]{www.jkkrishnan.in}

\credit{Conceptualization of this study, Methodology, Software,  Writing - Original draft preparation }

\affiliation[1]{organization={Sorbonne Université and CNRS, Institut Jean Le Rond d’Alembert UMR 7190},
                %addressline={4, Place Jussieu}, 
                city={F-75005 Paris},
%               citysep={}, % Uncomment if no comma needed between city and postcode
                %postcode={}, 
                %state={},
                country={France}}

\author[1,3]{Tian LONG}[orcid=0009-0006-2004-255X]
%\cormark[1]
\ead{tian.long@xjtu.edu.cn}
%ead[URL]{https://www.university.org}
\credit{Methodology, Software}

\author[1]{Jacob MAAREK}
%\fnmark[2]
\ead{jacob.maarek@sorbonne-universite.fr}
%ead[URL]{https://www.university.org}

\credit{Methodology, Software}

\affiliation[2]{organization={Institut Universitaire de France},
                %addressline={4, Place Jussieu}, 
                city={Paris},
%               citysep={}, % Uncomment if no comma needed between city and postcode
                %postcode={}, 
                %state={},
                country={France}}

\affiliation[3]{organization={State Key Laboratory for Strength and Vibration of Mechanical Structures, School of Aerospace, Xi'an Jiaotong University},
                %addressline={Street 29}, 
                postcode={710049}, 
                %postcodesep={}, 
                city={Xi'an},
                country={PR China}}

\author[1,2]{Stéphane ZALESKI}[orcid=0000-0003-2004-9090]
%\cormark[1]
%\fnmark[1,3]
\ead{stephane.zaleski@sorbonne-universite.fr}
\ead[URL]{http://www.lmm.jussieu.fr/~zaleski/}

\cortext[cor1]{Corresponding author}
%\cortext[cor2]{Principal corresponding author}

\begin{abstract}
  Three-dimensional direct numerical simulation of electrolysis is applied to investigate the growth and detachment of bubbles at electrodes. 
  The moving gas-liquid interface is modeled employing the VOF-based method. To ensure the accuracy of the simulations, 
  a mesh-independence study has been performed.
  The simulations include the growth phase of the bubbles, followed by their detachment from the electrode surface, 
  and the results are validated with analytical models and experimental data.
  The bubble growth is diffusion-controlled, leading to the scaling \(R\propto t^{1/2}\), but our simulation overpredicts the growth exponent during the initial stage.
  We further demonstrate that the number of nucleation sites significantly affects gas transport, as quantified by the Sherwood number. 
  The influences of contact angle and nucleation site on bubble detachment are also examined. 
  The predicted detachment radius varies linearly with contact angle, consistent with Fritz's linear relation
  between the volume-equivalent radius and contact angle, confirming that the surface tension is the dominant attachment force.
  Finally, as the nucleation sites increase, the induced bubble coalescence accelerates the bubble detachment. Taken together, 
  these findings give us valuable insights into improving gas bubble removal and enhancing overall electrolysis efficiency.
%\noindent\texttt{\textbackslash begin{abstract}} \dots 
%\texttt{\textbackslash end{abstract}} and
%\verb+\begin{keyword}+ \verb+...+ \verb+\end{keyword}+ 
%which
%contain the abstract and keywords respectively. 
%Each keyword shall be separated by a \verb+\sep+ command.
\end{abstract}

%\begin{graphicalabstract}
%\includegraphics{figs/cas-grabs.pdf}
%\end{graphicalabstract}

%\begin{highlights}
%\item Research highlights item 1
%\item Research highlights item 2
%\item Research highlights item 3
%\end{highlights}

\begin{keywords}
  multi-phase flows \sep phase change \sep electrolysis \sep bubble dynamics
\end{keywords}
\maketitle
\section{Introduction}\label{introduction}
The production of green hydrogen through water electrolysis is expected to be an important technology in achieving global net-zero emissions \citep{Turner2004, Holladay2009, Dawood2020}. However, it is slow and inefficient in many situations. 
The attached bubbles in such electrochemical devices reduce the efficiency of electrolyzer systems by blocking the active electrode sites or by increasing the ohmic resistance \citep{Swiegers2021}. 
Maintaining a bubble-free electrode surface is therefore crucial for highly efficient \(\mathrm{H_2}\) production.
Consequently, a detailed understanding of bubble evolution dynamics is highly desirable for developing strategies to enhance water electrolysis efficiency.
Analytical solutions for this problem are not generally available, and only a few exact solutions can be derived using extremely simplified assumptions \citep{Epstein1950, Scriven1959}.
However, they can still serve as valuable references for experimental and numerical studies \citep{Glas1964, Brandon1985, dapkus1986, vanderlinde2017a, Taqieddin2017, Soto2018, Angulo2020, Zhang2023}. 

Several experiments have confirmed that bubble growth is primarily governed by mass transport arising from gradients in the dissolved hydrogen concentration. 
\cite{Glas1964} performed a fundamental experiment for hydrogen bubbles generated on a flat electrode. 
The experimental findings demonstrate that the asymptotic growth of electrochemically generated bubbles follows the same functional relationship 
given by the analytical solution for a suspended bubble growing in a supersaturated liquid, see \cite{Scriven1959}. 
Recent studies have already investigated this problem on different scales. 
Macroscopic processes are mainly convection phenomena in the electrolyte solution, which is observed at the electrolytic cell scale.
\cite{li2018} visualized experimentally on two-phase flow at the anode side of a proton exchange membrane electrolyzer. 
They concluded that the inlet velocity does not affect bubble growth when the temperature and current density are constant. 
Furthermore, they examined hydrogen and oxygen bubble dynamics in a single-channel electrolyzer \citep{li2019}, 
revealing that the bubble detachment diameter varies inversely with flow velocity. Given the significant influence of electrolyzer geometry on macroscale bubbly flow, \citet{hreiz2015} analyzed a variety of electrochemical configurations.

 \cite{Abdelouahed2014} conducted Laser Doppler Velocimetry (LDV) measurements to observe the behavior of the bubble curtain. 
Other experimental studies revealed that the bubble curtain velocity is a function of the average electric current \citep{hine1980}, 
the mass transfer coefficient is varied for horizontal and vertical electrodes \citep{fouad1972}, 
and bubble coverage is also a function of current density \citep{vogt2005}. 
\cite{vogt2012} has mentioned that when the bubble coverage approaches unity, the reaction is totally blocked.
For the micro-scale, the micro-physical phenomena surrounding individual bubbles govern the overall behavior of the electrolysis process, 
and consequently, understanding the micro-scale bubble hydrodynamics is vital to reveal the governing mechanisms behind electrolysis. 
As the studied scale becomes smaller, however, the difficulty of the experimentation increases dramatically.
Several studies confirm that flow convection in the micro-area is caused by bubble growth, detachment, and coalescence. Such micro-convection
strongly influences the hydrogen diffusion boundary layer and changes mass transfer \citep{Stephan1979, vogt2005}. 
For example, \cite{Linde2018} used a specially designed electrode with pillars and pits to control micro-bubble formation, measuring the bubble radius and attachment force to verify the Fritz radius. 
Apart from bubble formation and growth, \cite{Bashkatov2024} investigated two modes of coalescence-induced bubble detachment using a dual platinum microelectrode system. 
By combining high-speed imaging and electrochemical analysis, they demonstrated the importance of bubble-bubble interactions in the departure process. 
The impact of micro-convection induced by bubble dynamics on mass transfer is significant. 
\cite{Burdyny2017} demonstrated that micro-convection enhances mass transport primarily by reducing the departure diameter of bubbles from the electrode surface. 
With advanced visualization techniques, the gas concentration released in the liquid by bubbles can be investigated \citep{dani2007, francois2011}. 
However, experimental measurements are generally expensive and constrained by the available measuring techniques, 
which usually provide global quantities (e.g., global mass transfer rate, rising velocities), and do not give information about local details, such as small-scale multiphase flow field and local mass transfer rate.

Due to the presence of large amounts of gas, conventional optical techniques are unable to detect many critical aspects of multiphase flow fields. 
Computational Fluid Dynamics (CFD) has become an important tool for the comprehensive study of electrolytic complex multiphase flows \citep{Hawkes2009, El-Askary2015}.
The significant increase in computational power has made direct numerical simulation an important alternative method for studying the detailed dynamics of mass transfer between fixed or deformable interfaces.

The individual bubble growth, multiple bubbles coalescence, 
and detachment that occur in this region have been numerically studied for a long period. It is confirmed that the diffusion-driven bubble growth dominates the growth stage \citep{Soto2018}. 
The bubble detachment radius is approximated by the Fritz radius \citep{vogt2004}. It can be derived from the force balance acting on the growing and detaching bubbles. 
Besides, the mass transfer mechanisms underlying this phenomenon require further discussion. \cite{vanderlinde2017a, Linde2018} simulated the hydrogen concentration field around a growing hydrogen bubble in acidic electrolysis, using a body-fitted axisymmetric finite difference method. 
\cite{sepahi2022a} employed an Immersed Boundary Method (IBM) to simulate the mass transport for bubbles in gas-evolving electrolysis, 
indicating that the net transport within the system is governed by the effective buoyancy driving induced by the rising bubble.
Very few simulations of deformable-bubble growth, detachment, and rise in the context of electrolysis have been performed using modern sharp-interface methods such as VOF, 
level set, or front tracking. Moreover, some authors used VOF methods without mass transfer \citep{lafmejani2017}. 
Most previous studies, such as those by \cite{sepahi2022a} and \cite{khalighi2023}, 
assumed undeformable spherical bubbles, neglecting bubble deformation that may arise due to lateral flow during detachment. 
Due to the time step limitations imposed by capillary effects, employing a rigid sphere approximation considerably reduces CPU time. 
Although these models are not suitable for designing actual systems, they still provide useful information and reference solutions for validating novel theories. Thus, simulations of bubble growth in electrolysis without the spherical bubble assumption have rarely been performed.
Recently, \cite{Gennari2022} proposed a VOF-based phase-change model for diffusion-driven mass transfer problems using the one-fluid method and a novel algorithm to extrapolate the discontinuous velocity field across the interface to improve interface advection accuracy. 
This method is applied to study the growth of deformable bubbles on planar electrodes. 

This study employs direct numerical simulation (DNS) to investigate all the aforementioned effects,
meaning that we have fully resolved all relevant scales of the hydrodynamic and concentration boundary layers. 
Compared to existing numerical studies, we are the first to utilize three-dimensional DNS to simulate the deformable interface motion of hydrogen bubbles at the electrode.
The fully resolved flow field provides crucial insight into the mechanisms of bubble growth, interaction, and detachment. All simulations are performed using the free code repository Basilisk (http://basilisk.fr/).

\section{Configuration and numerical methods}
\begin{figure}[htbp]
    \centering
    \includegraphics[width=0.9\linewidth]{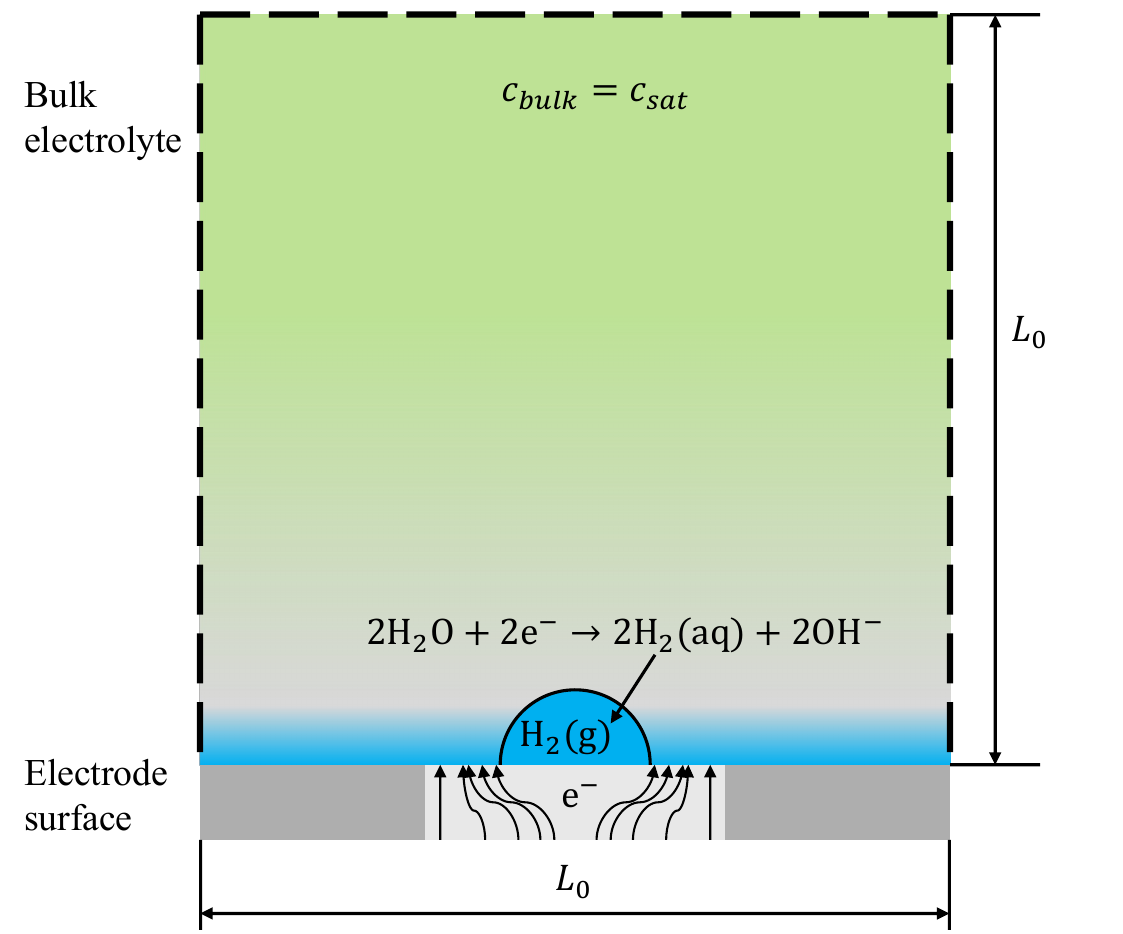}
    \caption{Schematic representation of the two-phase electrochemical system with relevant chemical reactions and boundary conditions at the cathode.}
    \label{fig: reaction}
\end{figure}

To simplify the complicated mathematical description of a bubble growing at an electrode in the presence of a surrounding flow, 
consider a two-phase gas-liquid system represented in \cref{fig: reaction}. 
We assume a constant room temperature, dilute liquid solutions ($\mathrm{KOH}$, $0.5 \mathrm{mol/L}$), 
and no evaporation of water. Besides, in order to avoid further complications, self-ionization of water is disregarded 
due to its low equilibrium constant at room temperature. The equation for the electric potential remains relatively simple. 
At the cathode, the reaction process is as follows,
\begin{align}
    \mathrm{2H_2O} + \mathrm{2e^-} \rightarrow  \mathrm{2H_2+2OH^-}.
\end{align}
In the present work, only the reaction at the cathode is considered.
\subsection{Problem set-up}\label{config}
\begin{figure}[htbp]
    \centering
    \includegraphics[width=0.9\linewidth]{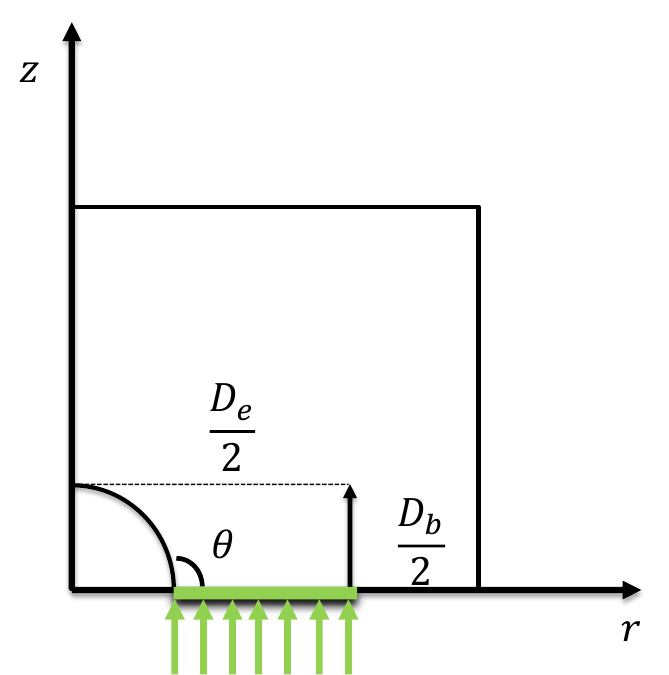}
    \caption{Sketch of the axisymmetric simulation setup.}
    \label{fig: ele_set}
\end{figure}

We start with axisymmetric simulations to validate the numerical mass transfer method provided by \cite{Gennari2022}.
Due to the geometric axisymmetric nature of the single bubble attached to a circular electrode, axisymmetric modeling not only effectively resolves the problem but also reduces computational costs while achieving a three-dimensional simulation.
A sketch of the axisymmetric setup and mesh grid is shown in \cref{fig: ele_set}. 
The symmetry axis is along $z$, while an outflow boundary condition is set on the top boundary.
The other boundaries are treated as no-slip walls. The computational domain is a square.  %Turning to the boundary conditions for 3D simulation, wall boundary conditions are employed for electrolysis cell sides (XZ planes and YZ planes ). At the outlet top boundary (top XY plane ), zero Neuman conditions are used for velocity and the pressure is set to zero. As for the electrode at the bottom or left side (vertical electrode)
The domain's size is \(L_0=25D_b \), where $D_b$ is the initial bubble diameter. The bubble is initialized in simulations with the diameter \(D_b=0.0127\) mm, 
The electrode is the flat end of a wire with a diameter of \(D_e=10 D_b=0.127\) mm, oriented along the radial axis (r axis).

\begin{figure}[htbp]
        \centering
        \includegraphics[width=0.9\linewidth]{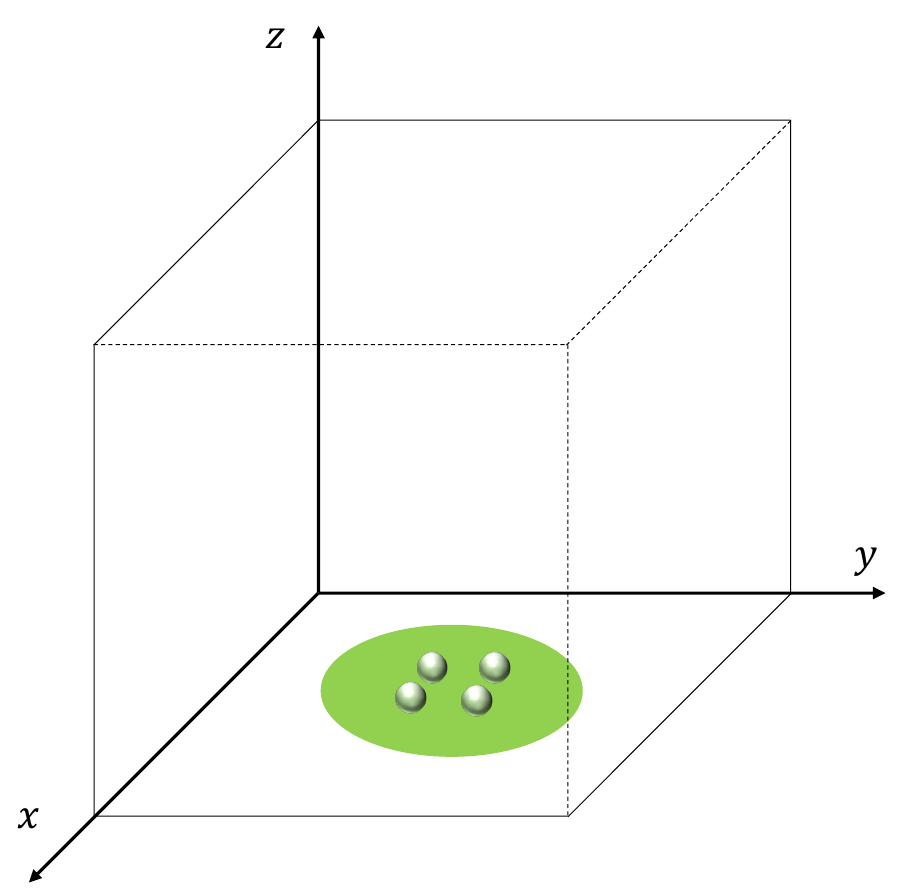}
        \caption{Sketch of the 3D simulation setup}
        \label{fig: sketch_3D}       
\end{figure}

%we investigate the variation of nucleation sites by comparing the results obtained for a single bubble with two bubbles 
%and four bubbles growing simultaneously. 
As for the more complex case of {\em multiple nucleation sites}, axial symmetry no longer holds. 
To address the mutual interaction between bubbles, a 3D configuration is required. The multiple nucleation sites are equally spaced at the electrode center,
as illustrated in \cref{fig: sketch_3D}.

The initial liquid is set to be saturated, and the saturation ratio is \(\zeta=c_0/c_s=1\), 
where \(c_s\) is the hydrogen concentration in the saturated liquid and \(c_0\) is 
the initial concentration of dissolved gas near the electrode surface. As mentioned in \cref{introduction}, the production of dissolved gas at the electrode walls creates a locally supersaturated 
region (\(\zeta>1\)). It drives the growth of bubbles formed from microscopic pits on the electrode surface due to the heterogeneous nucleation, see \cite{Jones1999,vanderlinde2017a}. 
According to the experiment, there should be no bubble present at the initial time \(t=0\). 
However, using the VOF method requires the volume fraction of gas to be initialized. 
We wait for a nucleation time before computing the volume change. 
The bubble size is fixed during this stage (\(t<t_n\)), while the concentration of dissolved hydrogen continues to increase. 
This approach enables the development of a concentration field around the bubble by the time nucleation occurs, better reflecting experimental observation.
The nucleation time varies due to many factors, like the electrode material and the current density. 
In the present work, we set the nucleation time \(t_n=0.02 \, \mathrm{s}\), a value derived from experimental work\citep{Glas1964}.
The control parameters for the electrolytically generated bubbly flow are the cathodic current density 
and the contact angles for different wettability of the electrode surface.
The current density \(I\) can give the molar flux of hydrogen (\(\mathrm{H_2}\)) by Faraday’s law,
\begin{align}
    J = \frac{I}{2F},
\end{align}
where \(I=i/A\), \(i\) is the total electric current,  \(A\) is the cross-section area (\(A=\pi/4 D_e^2\)); 
\(F\) is Faraday’s constant (\(F = 96485.3\) As/mol). 
To account for the flux of H\(_2\) across the active area of the electrode, 
a Neumann boundary condition for the gas concentration is applied to the electrode wall (r-axis for axis-symmetric simulation, 
xy plane for three-dimensional simulation),
\begin{align}
    \frac{\partial c}{\partial z} &= \frac{J}{D}  \qquad \text{for } r < \frac{D_e}{2},
    \\
    \frac{\partial c}{\partial z} &= 0    \qquad \text{for } r > \frac{D_e}{2}.
    \label{eq: electrode bc}
\end{align}
The boundary influx from the bubble-free region directly leads to an increase in the local concentration of dissolved hydrogen,
leading to the development of a hydrogen concentration layer. This process facilitates hydrogen bubble formation, 
driving its growth and detachment. Simulations performed in the present work are classified into two groups. The configurations are listed in \cref{tab: config}.
The axisymmetric configuration is primarily employed for numerical verification and sensitivity analysis, whereas the 3D configuration is used for most of the simulations on bubble growth evolution.
%The length of the cubic domain remains the same as the 2D-axis symmetry simulation (\(L_0=25D_b\)) and the finest mesh size (LEVEL 9) is also \(\Delta=L_0/2^9\). 

\begin{table*}[htbp]
    \centering
    \begin{tabular}{c c c c c}
        \hline
        \multicolumn{4}{c}{\textbf{Configuration}} \\
        \cline{1-4}
        \multicolumn{4}{c}{\textbf{Single bubble (Axisymmetric)}} \\
        \cline{1-4}
        \textbf{No.} & \(\theta\degree\) & \(z \times r (D_b)\) & mesh level\\
        \cline{1-4}
        1 & 35–90 (sweep: 35, 40, …, 90) & \(25\times 25\) & 8\\
        2 & 90(hydrophobic) & \(25\times 25\) & 9 \\
        3 & 90(hydrophobic) & \(25\times 25\) & 10 \\
        4 & 35(hydrophilic) & \(25\times 25\) & 9 \\
        5 & 35(hydrophilic) & \(25\times 25\) & 10 \\
        \cline{1-4}
        \multicolumn{4}{c}{\textbf{Three-dimensional}} \\
        \cline{1-4}
        \textbf{No.} & \(\theta\degree\) & \(x \times y \times z (D_b)\) & nucleation sites\\
        \cline{1-4}
        6 & 35–90 (sweep: 35, 40, …, 90) & \(25\times 25 \times 25\) & 1\\
        7 & 90(hydrophobic) & \(25\times 25 \times 25\) & 2 \\
        8 & 90(hydrophobic) & \(25\times 25 \times 25\) & 4 \\
        9 & 35(hydrophilic) & \(25\times 25 \times 25\) & 2 \\
        10 & 35(hydrophilic) & \(25\times 25 \times 25\) & 4 \\
    \end{tabular}
    \caption{Configurations for the three-dimensional and axisymmetric simulations.}
    \label{tab: config}
\end{table*}
The physical parameters remain the same for all the simulation cases (see \cref{tab: H2}).
Gravity acceleration is applied in the $-z$ direction with the value of \(9.8 \mathrm{m/s^2}\).  
%The numerics are limited by the very small value of the time step interval $\delta t$ and the massive number of time steps $N_T=T/\delta t$.To save the computation cost, the finest mesh size is \(Level=8\) based on the convergence study(AMR is used), which corresponds to 20 points\(/D_b\).

\begin{table}[htbp]
    \centering
    \begin{tabular}{c p{2.5cm} c c}
    \hline
    \textbf{Symbol} & \textbf{Properties} & \textbf{Value} & \textbf{Unit} \\
    \hline
    $\rho_c$ & Electrolyte density & 996 & [$\mathrm{kg/m^3}$] \\ 
    $\rho_d$ & Hydrogen density & 0.8 & [$\mathrm{kg/m^3}$] \\ 
    $g$ & Gravity acceleration & 9.8 & [$\mathrm{m/s^2}$] \\
    $M$ & Molar mass of hydrogen & 0.02 & [$\mathrm{kg/mol}$] \\ 
    $c_0$ & H$_2$ initial concentration & 0.002 & [$\mathrm{mol/m^3}$] \\ 
    $c_s$ & H$_2$ saturated concentration & 0.02 & [$\mathrm{mol/m^3}$] \\ 
    $\mu_c$ & Electrolyte viscosity & $8.32 \times 10^{-4}$ & [$\mathrm{kg/(m \cdot s)}$] \\
    $\mu_d$ & Hydrogen viscosity & $8.96 \times 10^{-6}$ & [$\mathrm{kg/(m \cdot s)}$] \\
    $\nu_c$ & Electrolyte kinematic viscosity & $8.35 \times 10^{-7}$ & [$\mathrm{m^2/s}$] \\   
    $\sigma$ & Surface tension & $7.5\times10^{-6}$ & [$\mathrm{N/m}$] \\
    $F$ & Faraday's constant & $96485.3$ & [$\mathrm{C/mol}$] \\
    $D$ & Hydrogen diffusion coefficient & $7.38 \times 10^{-9}$ & [$\mathrm{m^2/s}$] \\
    $D_b$ & Initial bubble diameter & $1.27 \times 10^{-5}$ & [$\mathrm{m}$] \\ 
    $R_0$ & Initial bubble radius & $6.35 \times 10^{-6}$ & [$\mathrm{m}$] \\ 
    $D_e$ & Electrode diameter & $1.27 \times 10^{-4}$ & [$\mathrm{m}$] \\   
    $t_n$ & Bubble nucleation time & $0.02$ & [$\mathrm{s}$] \\ 
    $P_0$ & Ambient pressure & $1.01325\times 10^{5}$ & [$\mathrm{N/m^2}$] \\
    $T_0$ & Ambient temperature & $298.15$ & [$\mathrm{K}$] \\
    $\mathcal{R}$ & Universal gas constant & $8.3145$ & [$\mathrm{J\cdotp mol^{-1} \cdotp K^{-1}}$] \\
    $H_s^{cp}$ & Henry's law solubility constant for $\mathrm{H_2}$ & $7.5\times10^{-6}$ & [$\mathrm{mol/(m^3Pa)}$] \\
    $P$ & Henry's law partition coefficient &53.3 &- \\
    \hline
    \end{tabular}
    \caption{Physical properties in SI units.}
    \label{tab: H2}
\end{table}

\subsection{Non-dimensional numbers}
The most basic non-dimensional number related to the transport of \(\mathrm{H_2}\) is the Sherwood number. The production rate of hydrogen flux is constant 
at the electrode, and \(J\) is a constant value in time. The transport of hydrogen results in a surface-averaged concentration, \(\bar{c_e}\) at the electrode surface.
To illustrate the transport, we compare \(\bar{c_e}\) with the bulk liquid concentration, defined here as the concentration at the top of the domain. Given that the electrolyte solution is initially saturated (\(\zeta=1\)), and \(c_0=c_s\), this yields the following Sherwood number for hydrogen production,
\begin{align}
    \overline{\mathrm{Sh_{e}}}=\frac{J D_b}{D(\bar{c_e}-c_s)}.
\end{align}
The bar symbol is used to mark the surface-averaged response parameters.
Introducing 
the boundary layer thickness \(\delta_\mathrm{H_2}=D(\bar{c_e}-c_s) /J\), 
this Sherwood number can be expressed as \(\overline{\mathrm{Sh_e}}=D_b/\delta_\mathrm{H_2}\). 
The boundary layer thickness is, in principle, time dependent, so the Sherwood number also is time dependent. However, after several bubble detachments or in the presence of externally forced flow, the boundary layer thickness would eventually reach a statistical steady state, and a statistically steady Sherwood number will be reached. 

For mass transfer accounting for local volume change, the Sherwood number of the bubble is expressed as follows to quantify the hydrogen transport across the interface,
\begin{align}
    \overline{\mathrm{Sh_{b}}}=\frac{D_b \int_{\Sigma} \dot{m}\,ds}{A_{\Sigma} M D(\bar{c_e}-c_s)},
\end{align}
where \(A_{\Sigma}=4\pi R^2\) is the interface surface area for a spherical bubble. And \(\bar{c_e}-c_s\) represents the concentration difference between the electrode and bubble interface as the concentration of the interface \(c_{\Sigma}=c_s\). Since we assume the configuration with an ambient temperature and pressure, the mass transfer rate across the interface can be defined as
\begin{align}
    \int_{\Sigma} \dot{m}\,ds = \left(\frac{P_0}{\mathcal{R} T_0}\right) 4\pi R^2 \frac{dR}{dt} M,
\end{align}
where \(\mathcal{R}, T_0, P_0\) are the universal gas constant, ambient temperature, and pressure, respectively. Then it leads to a rather simple expression of the bubble Sherwood number,
\begin{align}
    \overline{\mathrm{Sh_{b}}}=\frac{P_0}{\mathcal{R} T_0} \frac{2R}{D(\bar{c_e}-c_{\Sigma})}\frac{dR}{dt}.
    \label{eq: sh_b}
\end{align}

Apart from the Sherwood number, which measures the mass transfer, other important nondimensional numbers are the Schmidt,  Galileo, and Bond numbers.
First, the Schmidt number compares momentum diffusion to mass diffusion and is defined as
\begin{align}
    \mathrm{Sc}=\frac{\nu_c}{D},
\end{align}
where the \(\nu_c\) is the kinematic viscosity of the electrolyte.
The ratios of dominant forces acting on the bubble can be described using two additional dimensionless numbers: the Galileo and Bond numbers. The Galileo number, which compares gravitational to viscous forces, is given by
\begin{align}
    \mathrm{Ga}=\sqrt{\frac{\rho_cgD_b^2}{\nu_c^2}}.
\end{align}
When the bubble approaches detachment, gravitational effects become dominant and can be quantified by the Bond number, which represents the ratio of gravitational to capillary forces
\begin{align}
    \mathrm{Bo}=\frac{\rho_cgD_b^2}{\sigma}.
\end{align}

\subsection{Fluid dynamical equations}\label{sec: governing}
A model of the microscopic process involves solving the two-phase incompressible Navier-Stokes equations with phase change, surface tension, gravity, and contact line dynamics on the wall.
Due to the constant temperature assumption, no equation for the thermal energy is needed. 

The gas phase is called $\Omega_d(t)$, for the disperse phase, and the liquid phase is called $\Omega_c(t)$, for the continuous phase. 
These two subdomains are separated by an infinitely thin interface $\Sigma(t)$.
The entire domain is given by $\Omega=\Omega_d(t) \cup \Omega_c(t) \cup \Sigma(t)$.
The normal vector $n_\Sigma$ at the interface points into $\Omega_d(t)$.

For each phase, the governing equation of an incompressible flow system in the absence of mass transfer reads,
\begin{align}
  \nabla \cdot \mathbf{u} &= 0 \quad \text{in } \Omega \backslash \Sigma,  \label{eq: continuity} \\
  \begin{split}
      \partial_t (\rho \mathbf{u}) + \nabla \cdot (\rho \mathbf{u} \otimes \mathbf{u}) 
      &= -\nabla p + \nabla \cdot (2\mu \mathbf{D}) \\
      &\quad + \rho \mathbf{g} \quad \text{in } \Omega \backslash \Sigma,
  \end{split} \label{eq: momentum}
\end{align}
where the density $\rho$ and viscosity $\mu$ remain constant in $\Omega_d$ and $\Omega_c$.
The \cref{eq: continuity} is the continuity equation, where $\mathbf{u}$ represents the velocity field. 
In the balance of momentum \cref{eq: momentum}, $p$ is the static pressure, $\mathbf{D}$ is the deformation tensor,
and $g$ represents body force, which is the gravitational acceleration in this system. So, $a$ is replaced by $g$ in the following sections.
The \cref{eq: continuity} and \cref{eq: momentum} are valid everywhere in the domain except at the interface, 
where additional conditions are needed \citep{Tryggvason_Scardovelli_Zaleski_2011}. 
The continuity equation requires that the amount of mass that leaves one phase $\Omega_d (\Omega_c)$ 
must be transferred to another phase $\Omega_c (\Omega_d)$ since the infinitely thin interface region can not store any mass.
It results in a jump condition across the interface,
\begin{align}
  \Vert\rho (\mathbf{u}-\mathbf{u_\Sigma})\cdot \mathbf{n_\Sigma}\Vert = \Vert \dot{m}\Vert= 0 \label{eq: jump1},
\end{align}
where the jump notation has been introduced(e.g. $\Vert\rho\Vert=\rho_c-\rho_d$); 
$\mathbf{u_\Sigma}$ is the interface velocity and $\dot{m}$ is the mass transfer rate $\mathrm{kg/m^2 s}$. 
The second jump condition is derived by applying the conservation of momentum to a control volume 
with an infinitely small thickness around the interface, and it reads

\begin{align}
    \Vert \rho\mathbf{u} \otimes (\mathbf{u}-\mathbf{u_\Sigma}) + p\mathbf{I} -2\mu \mathbf{D} \Vert \cdot \mathbf{n_\Sigma} = \sigma k \mathbf{n_\Sigma} + \nabla_\Sigma \sigma, \label{eq: jump2}
\end{align}
where $\mathbf{I}$ is the unit tensor, $\sigma$ is the surface tension and $k$ is the curvature of the interface. In the problem of hydrogen bubble growth, the interface is considered to have uniform surface tension; besides, a no-slip boundary condition is applied at the interface. \cref{eq: jump2} can be further simplified \cite{fleckenstein2015} in the form
\begin{align}
    \Vert p\mathbf{I} -2\mu \mathbf{D} \Vert \cdot \mathbf{n_\Sigma}= \sigma k \mathbf{n_\Sigma}. \label{eq: jump3}
\end{align}
The numerical method used in this work for interface transport is the Volume of Fluid (VOF) method, and is combined with a one-fluid formulation of the governing equations.
In the one-fluid approach, the jump conditions \cref{eq: jump1} and \cref{eq: jump3} are replaced by source terms that 
act at the interface as singularities ($\delta$ function), a single set of Navier-Stokes equations for the entire domain $\Omega$ 
is solved as follows
\begin{align}
    \nabla \cdot \mathbf{u} &= \dot{m}\left(\frac{1}{\rho_d}-\frac{1}{\rho_c}\right) \delta_\Sigma,  \label{eq: continuity_1} \\
    \partial_t \mathbf{u} + \nabla \cdot ( \mathbf{u} \otimes \mathbf{u}) &= \frac{1}{\rho}\left[-\nabla p + \nabla \cdot (2\mu \mathbf{D})\right] + \frac{\sigma k \mathbf{n_\Sigma}}{\rho}\delta_\Sigma, \label{eq: momentum_1}
\end{align}
where the $\delta_\Sigma$ is the surface Dirac distribution. 
Therefore, the system of \cref{eq: continuity_1,eq: momentum_1} is valid for the whole domain. 
Then, to determine the location of the interface, a marker function is required. The Heaviside function serves this purpose
\begin{align}
    H(x, t) =
\begin{cases} 
1, & \text{if } x \in \Omega_c , \\
0, & \text{if } x \in \Omega_d .
\end{cases}\label{eq: H_define}
\end{align}
%where the continuous phase is assumed to be the primary phase, i.e., where $H(x, t) = 1$. Once $H(x, t)$ is known everywhere, the values of $\rho$ and $\mu$ can be computed as:
%\begin{align}
%    \rho = \rho_c H + \rho_d(1 - H), \label{eq: rho}
%\end{align}
%and
%\begin{align}
%    \mu = \mu_c H + \mu_d(1 - H), \label{eq: mu}
%\end{align}
The transport equation for Heaviside function $H(x,t)$ can be obtained from the following integral balance for a control volume $V$
\begin{align}
  \begin{split}
      \int_V \partial_t H \, dV 
      + \oint_{\partial V} H \mathbf{u} \cdot \mathbf{n} \, dS \\
      + \int_\Sigma (\mathbf{u}_c - \mathbf{u}_\Sigma) \cdot \mathbf{n}_\Sigma \, dS 
      &= 0,
  \end{split} \label{eq: H_int}
  \end{align}  
where the second term on the LHS represents the convective transport, and the last term is a source term that accounts for the mass transfer across the interface, which is null when $\dot{m} = 0$. 
Converting the surface integral to a volume integral, we can write it in differential form,
\begin{align}
    \partial_t H + \nabla \cdot (H \mathbf{u}) + \frac{\dot{m}}{\rho_c} \delta_\Sigma = 0. \label{eq: h_trans}
\end{align}
\cref{eq: h_trans} is then used to compute the volume fraction, see \cref{Sec: solver}.

\subsection{Concentration transport equation}\label{transport}
This model can be used to compute the concentration field of the soluble hydrogen 
in a two-phase system with mass transfer by applying the two-scalar method of \cite{fleckenstein2015}. 
In the present study, we focus on pure incompressible gas bubbles, and we assume that 
no electrolyte species exists in the gas phase (Phase (i.e., the electrolyte is not volatile). 
The system contains two species: hydrogen (denoted by subscript 1) and electrolyte liquid (denoted by subscript 2). 
Overall mass transfer is entirely governed by the transport of the hydrogen species.
It is worth pointing out that we do not need to solve the mass balance 
in the dispersed phase since no mixture exists inside the bubbles (pure hydrogen gas).
The mass balance of hydrogen in the continuous phase domain $\Omega_c$ reads
\begin{align}
    \partial_t\rho^1+\nabla \cdot (\rho^1 \mathbf{u^1})=0,\label{eq: mass_balance}
\end{align}
where $\rho^1$ is the partial density of hydrogen and $\mathbf{u^1}$ denotes the hydrogen velocity. It is coupled with the jump condition for the conservation of mass
\begin{align}
    \Vert \rho^1 (\mathbf{u^1}-\mathbf{u_\Sigma})\cdot \mathbf{n_\Sigma}  \Vert = \Vert \dot{m}^1 \Vert =0. \label{eq: jump4}
\end{align}
The average phase density and velocity are derived from the respective species terms
\begin{align}
    \rho_c=\rho^1+\rho^2,
\end{align}
and 
\begin{align}
    \rho_c\mathbf{u}=\rho^1 \mathbf{u^1}+\rho^2 \mathbf{u^2},
\end{align}
where the superscript $2$ denotes the electrolyte species.    Then the transport equation of hydrogen for incompressible flow can be written as
\begin{align}
    \partial_t\rho^1+\mathbf{u}\cdot \nabla \rho^1 +\nabla \cdot J^1=0, \label{eq: mass_balance_1}
\end{align}
and the diffusive flux of hydrogen is
\begin{align}
    J^1=\rho^1(\mathbf{u^1}-\mathbf{u}).
\end{align}
The mass-transfer rate of hydrogen can be derived from \cref{eq: jump4,eq: jump1}
\begin{align}
  \begin{split}
      \dot{m}^1 
      &= \rho^1(\mathbf{u} - \mathbf{u}_\Sigma) \cdot \mathbf{n}_\Sigma 
      + \rho^1(\mathbf{u}^1 - \mathbf{u}) \cdot \mathbf{n}_\Sigma \\
      &= \frac{\rho^1}{\rho_c} \cdot \dot{m} + J^1 \cdot \mathbf{n}_\Sigma,
  \end{split} \label{eq: mass_trans_1}
\end{align}
which shows that the mass transfer contains both a convective term and a diffusive term. Under the assumption of dilute liquid solutions, the diffusive flux can be well modeled by Fick’s law of diffusion
\begin{align}
    J^1=-D^1 \nabla \rho^1, \label{eq: flux}
\end{align}
where the $D^1$ is the hydrogen diffusion coefficient. Combining Eqs (\ref{eq: flux}) and (\ref{eq: mass_trans_1}), the mass-transfer rate of hydrogen reads
\begin{align}
    \dot{m}=-\frac{M^1D^1}{1-\frac{\rho^1}{\rho_c}}\frac{\partial c^1}{\partial n_\Sigma}, \label{eq: mass_trans_2}
\end{align}
where the molar concentration has been introduced, i.e. $c^1=\rho^1/M^1$ with $M^1$ the molar mass. 
It provides the mass-transfer rate evaluated from the continuous side of the interface, which corresponds to $n_\Sigma=1$.
Since only hydrogen can be transferred between phases, we will omit the species indicator in what follows. We will only refer to the concentration of soluble hydrogen in the liquid phase domain $\Omega_c$.

To compute the concentration gradient, we need the hydrogen concentration at the liquid side of the interface. 
For a gas-liquid system at equilibrium, we can employ Henry’s law to compute the concentration on the liquid side of the interface
\begin{align}
    (c_c)_\Sigma=\frac{(c_d)_\Sigma}{P},
    \label{eq: henry}
\end{align}
where $P$ is the partition coefficient, which can be deduced from $P=1/({\mathcal{R}T_0H_s^{cp}})$, and is taken as a constant for the present work. 
So, the hydrogen concentration at the liquid side of the interface \((c_c)_{\Sigma}\) is immediately computed. 
$(c_d)_\Sigma = \rho_d/M$ is a constant, as the density should be constant everywhere inside the bubble ($\Omega_d$).

For numerical integration, we rewrite the mass balance \cref{eq: mass_balance_1} for the hydrogen diffusion process. 
As we discussed, the mass balance will be done in the continuous region $\Omega_c$ and for a non-reactive flow ($R=0$). 
Since the hydrogen dissolved in $\Omega_c$ is the only species that we need for mass transportation
\begin{align}
  \begin{split}
      \int_V \partial_t c \, dV 
      &+ \oint_{\partial V} (c \mathbf{u} - D \nabla c) \cdot \mathbf{n} \, dS \\
      &+ \int_\Sigma c (\mathbf{u} - \mathbf{u}_\Sigma) \cdot \mathbf{n}_\Sigma \, dS \\
      &= 0,
  \end{split}
  \end{align}  
The phase indicator will be omitted in the following, i.e., $c=c_c$.
Combine the jump condition \cref{eq: jump4}, and the final differential form could be deduced,
\begin{align}
    \partial_t c + \mathbf{u}\cdot\nabla c = \nabla\cdot\left(D\nabla c\right) - \frac{\dot{m}}{M}\delta_\Sigma.
    \label{eq: species transport equation}
\end{align}

\subsection{Numerical methodology}\label{Sec: solver}
The phase change model is derived from the work of \cite{Gennari2022}. 
The governing equation shown in \cref{sec: governing} is solved using the free open science platform Basilisk (HTTP://basilisk.fr/),
which provides finite-volume partial differential equation (P.D.E.) solvers on adaptive cartesian grids. 
Using quadtree/octree adaptive mesh refinement (AMR) in regions with large gradients makes the approach particularly suitable for multiscale 
processes such as interfacial flows. 
In interfacial flows, a fine mesh is typically required around the gas-liquid interface but not the entire domain. 
The shape of the domain is always a square $L_0 \times L_0$ in the axisymmetric configuration (a cube $L_0 \times L_0 \times L_0$ in three dimensions).
The grid is organized following a hierarchical quadtree/octree structure, 
where each cell can be further divided into four child cells (eight in three dimensions), 
and a level is assigned to each cell according to its position in the tree structure. 
The root cell is at level 0, and its size $\Delta$ is the same as that of the whole numerical domain ($\Delta = L_0$). 
A generic cell at level $l$ has size $\Delta(l) = \frac{L_0}{2^l}$. 
The grid structure in Basilisk allows neighboring cells to vary by up to one level, 
meaning each cell edge/face can communicate with no more than two finer edges/faces.
The VOF method is one of the most widely used numerical approaches for the modeling of two-phase immiscible fluids.
The starting point for the derivation of the VOF approach is the one-fluid formulation presented in \cref{sec: governing}, 
which is vital to the phase change problem. The volume fraction of the continuous phase is defined as
\begin{align}
f_c &= \frac{1}{V} \int_V H \, dV ,
\end{align}
and the value of $f_c$ is within the set $[0, 1]$, depending on the amount of liquid in the cell.
\begin{align}
f_c = 
\begin{cases} 
0 & \text{if the cell is pure gas}, \\
1 & \text{if the cell is pure liquid}, \\
[0,1] & \text{if the cell is mixed}.
\end{cases}
\end{align}
The volume fraction of the dispersed phase $f_d$ is implicitly described by the relationship $f_c+f_d=1$. 
So, only one transport equation of H needs to be solved. By applying the incompressibility constraint, 
the integrated form of \cref{eq: h_trans} is
\begin{align}
  \begin{split}
      \frac{\partial}{\partial t} \int_V H \, dV 
      &+ \frac{1}{V} \int_V \nabla \cdot (H \mathbf{u}) \, dV \\
      &+ \frac{1}{V} \int_V \frac{\dot{m}}{\rho_c} \delta_\Sigma \, dV 
      = 0.
  \end{split} \label{eq: h_int}
  \end{align} 
The integration of the transport \cref{eq: h_int} is performed in two steps, namely the reconstruction step and the propagation step. 
First, the interface is approximated with a line/plane in each interfacial cell. Second, the fluxes of volume fraction across the cell 
boundaries are computed, and \cref{eq: h_int} is integrated in time. The geometric reconstruction of the interface is based on the 
piecewise linear interface construction (PLIC) method, where the interface is approximated as a line (plane) in an axisymmetric (three-dimensional) configuration.

\section{Implementation}
In this section, we show the simulation results for several different configurations, as described in \cref{config}.
This numerical study is designed to simulate the experiments conducted by \cite{Glas1964}. 
The experimental results confirm that the driving force for bubble growth is diffusion and measure the bubble growth rate.
The physical properties corresponding to an alkaline solution (typically used in industrial water electrolysis) 
are shown in \cref{tab: H2}.

It should be noted that the original surface tension  \(\sigma=0.075\mathrm{N/m}\)  is decreased by a factor of \(10^{-4}\) and the hydrogen density \(\rho_d = 0.08 \, \text{kg/m}^3\) is increased to \(\rho_d = 0.8 \, \text{kg/m}^3\). The molar mass is scaled by the same factor, ensuring that the relative volume change remains unaffected (\(\Delta V \propto M/\rho_d\)). 
Sensitivity tests are therefore required to assess the effects of the decreased surface tension and increased hydrogen density.

\begin{figure*}[htbp]
    \begin{subfigure}[b]{1\textwidth}
        \centering
        \includegraphics[width=0.9\textwidth]{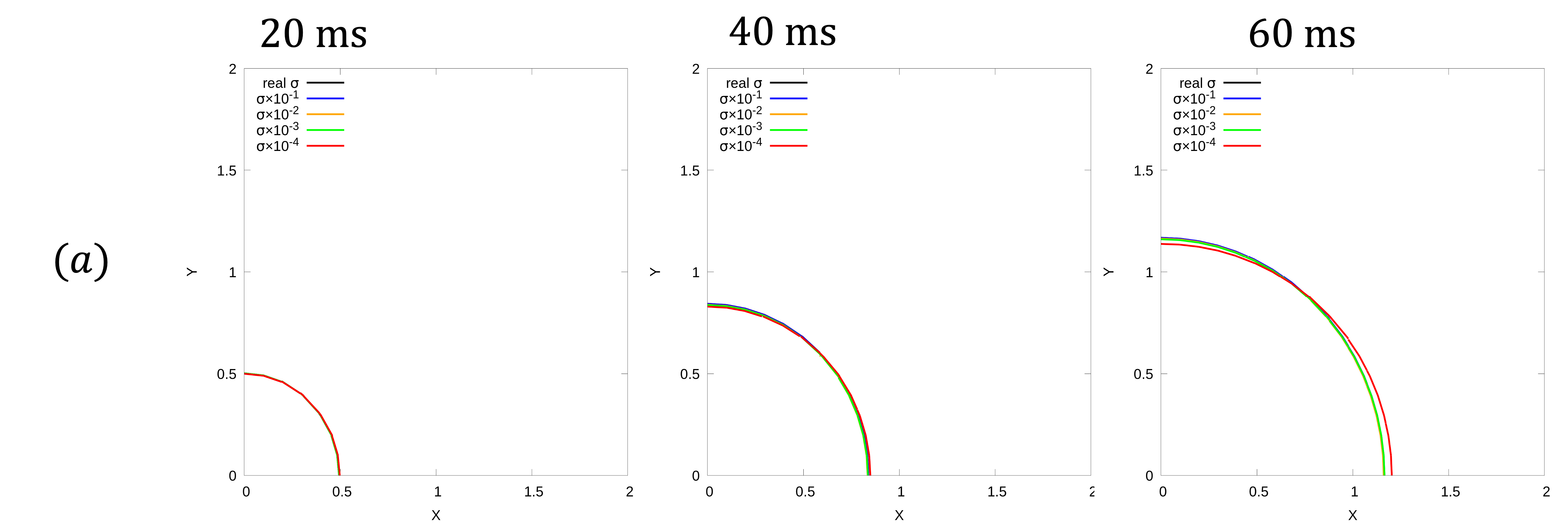}
        \label{fig: surface}
    \end{subfigure}
    
    \vspace{0.5cm}
    
    \begin{subfigure}[b]{1\textwidth}
        \centering
        \includegraphics[width=0.9\textwidth]{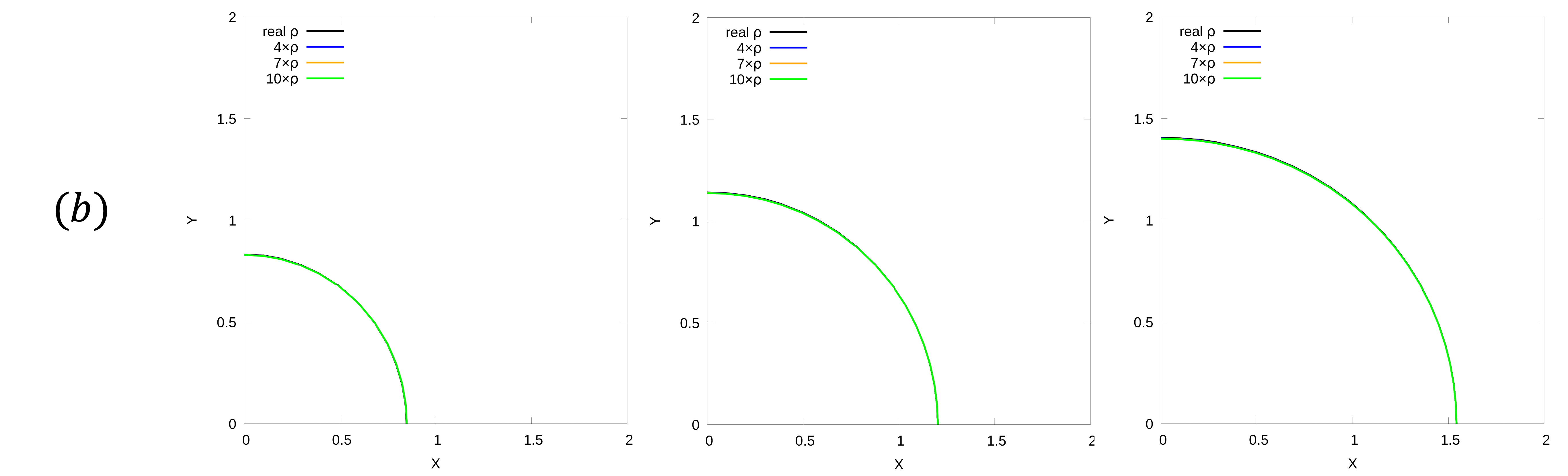}
        \label{fig: density}
    \end{subfigure}
    
    \caption{Sensitivity test on decreased surface tension (a), and increased hydrogen density (b).}
    \label{fig: vertical}
\end{figure*}

Decreasing the surface tension is introduced to reduce computational cost, as the surface-tension scheme imposes a restrictive time-step constraint. A comparison of simulations with modified \(\sigma\) in \cref{fig: vertical} indicates that increasing surface tension has only a minor influence on bubble growth. An analysis of the Weber number for different values of \(\sigma\) further shows that surface tension remains the dominant force during the bubble growth stage, even after \(\sigma\) is decreased. Bubble detachment, however, is significantly more sensitive to the choice of \(\sigma\), as reflected by the Bond number. To maintain consistency with the theoretical prediction of the detachment radius, the same decreased value of \(\sigma\) is adopted.

Regarding hydrogen density, its physical value is relatively small, leading to a large density ratio \((\rho_c / \rho_d = 12450)\), which would be a problem for numerical simulations. A high density ratio is observed to slow the convergence of the multigrid solver. Sensitivity tests for different hydrogen densities, shown in \cref{fig: vertical}, demonstrate that increasing \(\rho_d\) has a negligible effect on bubble growth. Based on the results above, we can confirm that a decrease of surface tension $\sigma$ by a factor of $10^{-4}$ and an increase of hydrogen density to $10\times \rho_d$ provide an acceptable balance between numerical stability, computational efficiency and physical accuracy.
The simulation results in this work are presented in dimensional units to facilitate comparison with experimental measurements.
According to the analytical and experimental work, the growth of electrochemically generated bubbles follows the same functional 
relationship as the solution for bubble growth in a supersaturated liquid described by the Scriven model.
Different stages or regimes for bubble growth have been characterized by a power law: growth controlled by inertia 
(\(R \propto t\)), by diffusion (\(R \propto t^{1/2}\)) or by reaction limitation (\(R \propto t^{1/3}\)) \citep{Angulo2020}. 
The growth behavior (\(R\propto t^{1/2}\)) reflects a standard analytical solution for diffusive bubble growth.
To validate the specific relationship, we first perform a series of axisymmetric simulations on bubble growing at the contact angle of \(\theta=90\degree\).
%Because the Scriven model assumes radially symmetric hemisphere growing on a flat plane, which corresponds to the contact angle of \(\theta=90\degree\).
\subsection{Axisymmetric simulation validation}\label{sec: 2D_verification}
\begin{figure}[htbp]
    \centering
    \includegraphics[width=0.9\linewidth]{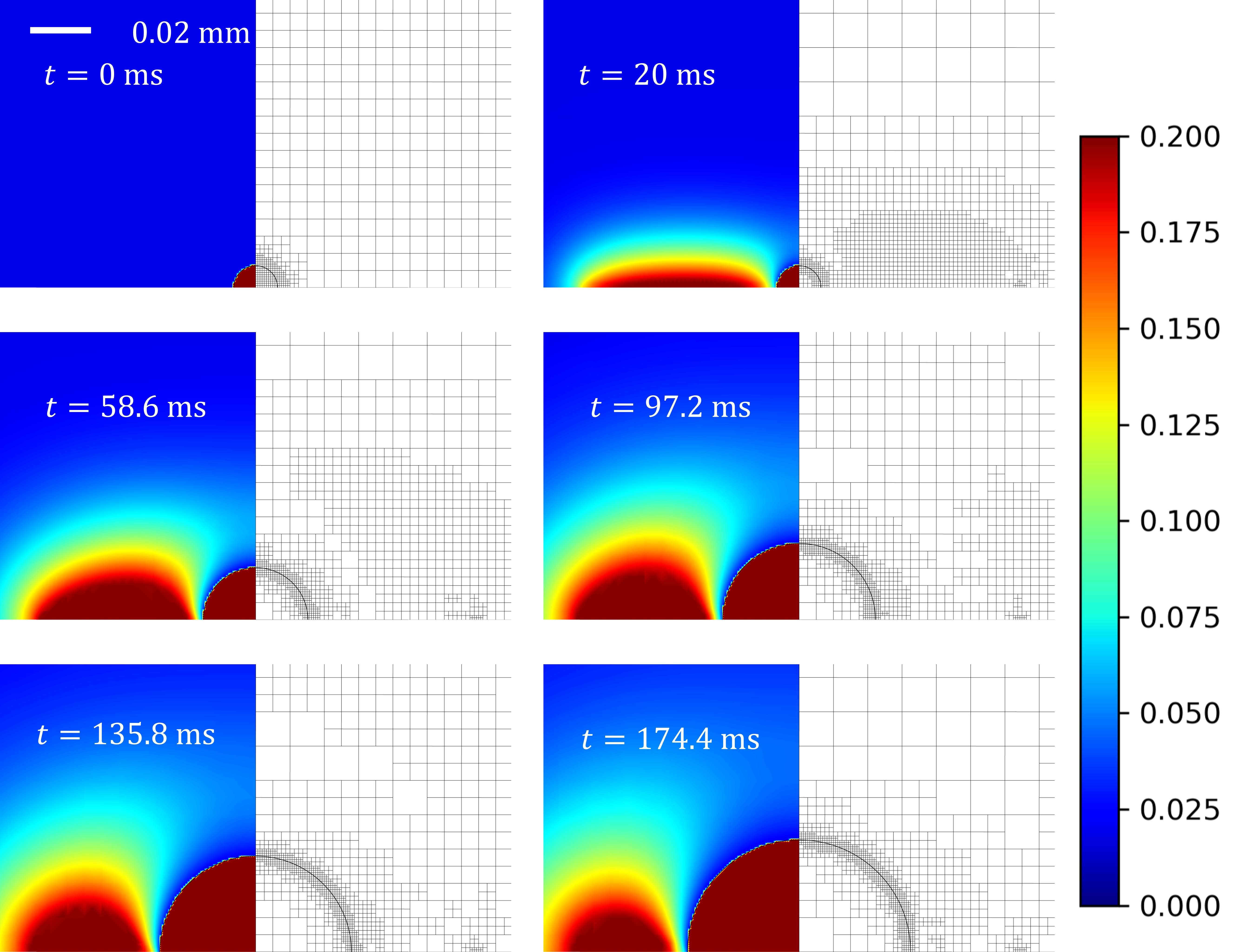}
    \caption{Frames show the dissolved hydrogen concentration for a bubble growing at the contact angle of \(\theta=90 \degree\)(with the current density of \(I=\mathrm{1000 A/m^2}\)). The color legend represents \(\mathrm{H_2}\) concentration, while the upper limit is $0.2 \mathrm{mol/m^3}$.}
    \label{fig: con}
\end{figure}

In this Section, we assume there is a single bubble (with an initial diameter \(D_b\)) on the cathode. In \cref{fig: con}, the left-hand side of the snapshots
illustrates hydrogen evolution during the bubble's growth stage, where color represents the dissolved hydrogen concentration in the liquid phase. 
For visualization consistency, the gas phase inside the bubble will also be colored, but will not be calculated (based on the assumption of constant pressure inside the bubble) and will remain at $1\mathrm{mol/m^3}$.
The same color map is applied to the rest of the snapshots in the following. Hydrogen is produced from the electrode surface, and the saturation near the electrode wall is more significant than that in the bulk liquid. 
Completing the visualization, the right side of the snapshots displays the grid and the moving interface, clearly showing adaptive quadtree mesh refinement.

\begin{figure}[htbp]
    \centering
    \includegraphics[width=0.9\linewidth]{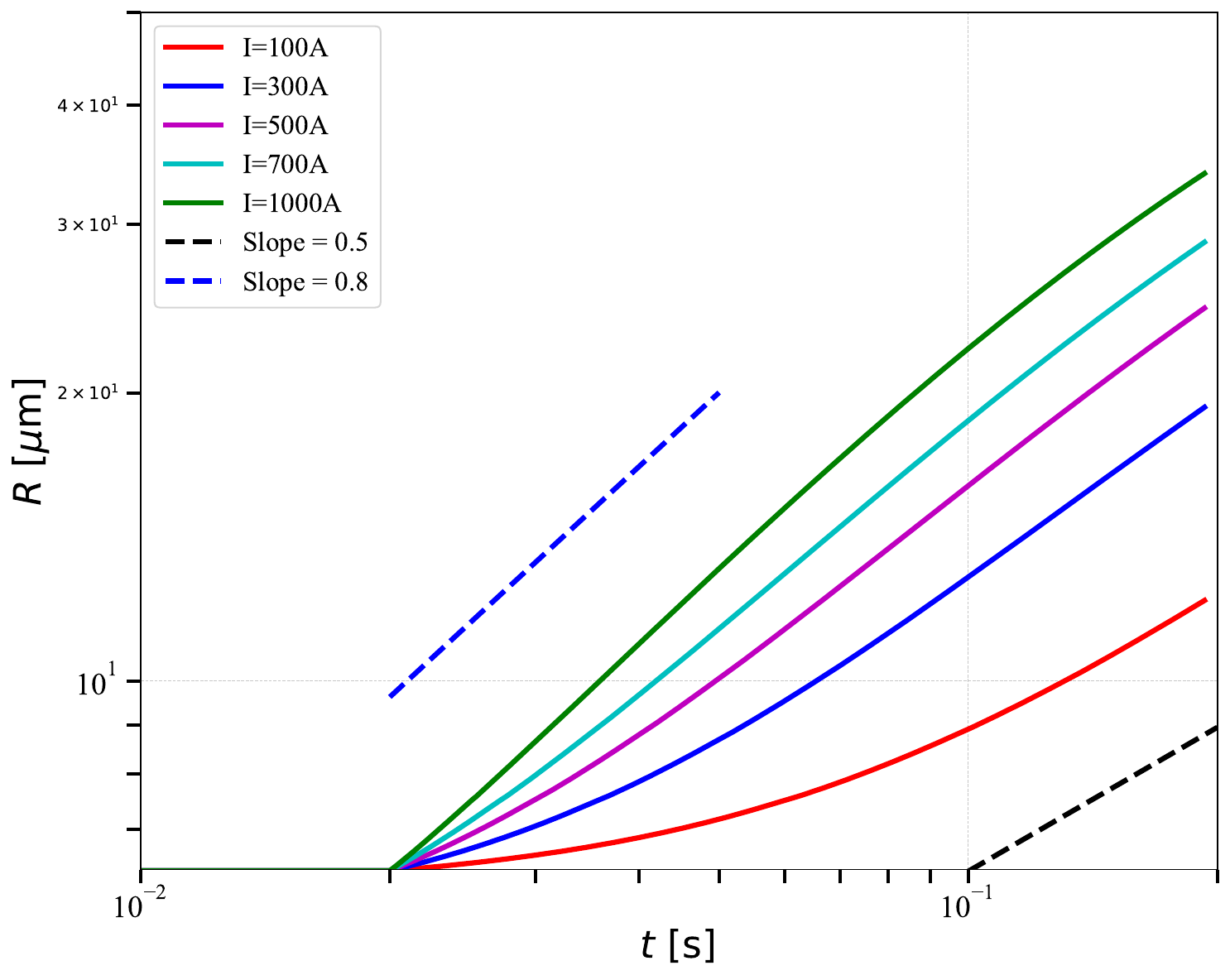}
    \caption{The bubble radius growth slope and growth exponent variation.}
    \label{fig: ele_log}
\end{figure}

A mesh independence study has been performed to demonstrate that a mesh refinement level of 8 will be sufficient for accurate resolution.
The following logarithmic plot \cref{fig: ele_log} presents bubble growth for five different current densities between \(t_n=0.02 \, \mathrm{s}\) and \(t=0.2 \, \mathrm{s}\). 
In our simulation case, the hydrogen flux is applied before the bubble nucleation time (\(t_n=20 \mathrm{ms}\)), 
with the average hydrogen concentration on the electrode wall increasing in time. After nucleation, the concentration difference drives bubble growth through the diffusion process.

The inertia-driven and diffusion-driven growth regimes correspond to slopes of 1 and \(1/2\), respectively. As shown in \cref{fig: ele_log}, the transition to diffusion-driven growth is evident, but the expected inertia-driven regime is missing. The steepest slope is observed only during the initial growth stage at the highest current density, and is close to $0.8$.

The diffusion-driven growth slope of \(1/2\) is assumed to be valid for a spherically symmetric concentration field, a bubble radius much larger than the initial radius, 
constant solubility, and, far from the bubble, constant concentration and zero velocity. This solution is confirmed in cases where the 
thickness \(\delta_{\mathrm{H_2}}\) of the diffusion boundary layer surrounding the bubble is small compared with the diameter of the bubbles. Given these constraints of the analytical solution, the discrepancies in our simulation can be explained. 
The hydrogen flux at the boundary, which is determined by current density, influences the growth exponent by increasing the averaged hydrogen concentration on the electrode wall. At the beginning stage, the growth exponent is larger than \(1/2\). The bubble radius is relatively small compared to the thickness of the diffusion boundary layer.

At high current densities, such as \( I = 1000 \, \mathrm{A/m^2} \), the increased hydrogen flux raises the supersaturation level and generates a strong diffusion boundary layer. This layer continuously supplies \(\mathrm{H_2}\) for bubble growth, resulting in a production rate that surpasses consumption and yields a growth exponent greater than \(1/2\). Numerically, the stronger flux produces a steeper \(\mathrm{H_2}\) concentration gradient, which enhances mass transfer across the interface. In contrast, lower current densities limit hydrogen availability, making it insufficient to sustain bubble growth and leading to sub-\(1/2\) exponents. The decreasing trend of the growth exponent with lower current densities confirms this behavior.

After the initial growth stage, as the bubble radius increases, the \(\delta_\mathrm{H_2}\) is much smaller compared with the bubble radius. 
The growth exponents for different boundary conditions are all approaching \(1/2\), which indicates that the overall growth is diffusion-controlled.

\subsection{Three-dimensional simulation}
To extend the numerical scheme to three-dimensional simulation, we set the domain as a cube (\cref{fig: sketch_3D}). 
And the boundary condition would be modified to suit the domain. The electrode will be placed in the middle of the domain bottom boundary ($xy$ plane), 
which will give the same boundary condition of hydrogen flux as an axisymmetric simulation, see \cref{eq: electrode bc}.
The top $xy$ plane will be set as an outflow boundary condition, and on the lateral boundaries ($xz$ and $yz$ planes), a no-slip boundary condition is set. 

A single bubble is modeled in three dimensions to validate the applicability of the numerical method. We compare the bubble growth prediction of the three-dimensional simulation to that of the corresponding axisymmetric simulation.
Specifically, simulations of configurations No. 1 and No. 6 (\cref{tab: config}) are compared. 
The results show that the bubble growth slopes obtained from the axisymmetric simulations are identical to those from the three-dimensional simulation. Since the numerical convergence has been verified in \cref{sec: 2D_verification}, this good agreement confirms the consistency of the three-dimensional simulation. Consequently, the three-dimensional configuration is used for the subsequent simulations.

\section{Result and discussion}
\subsection{Single bubble growth}\label{sec: single_growth}
Based on the experiments of \cite{Glas1964}, both the current density and the electrode surface covering (determined by the contact angle $\theta$) influence the bubble growth.
A notable observation is that the contact angle is variable. \cite{Glas1964} found that the change is large, from about 70\degree  to 20\degree. However, for simplicity, our numerical model assumes a constant contact angle throughout the bubble growth evolution.
For further comparison, another series of simulations is performed with a smaller contact angle (\(\theta = 35^\circ\)), which corresponds to a hydrophilic electrode surface. 

\begin{figure}[htbp]
    \centering
    \includegraphics[width=0.9\linewidth]{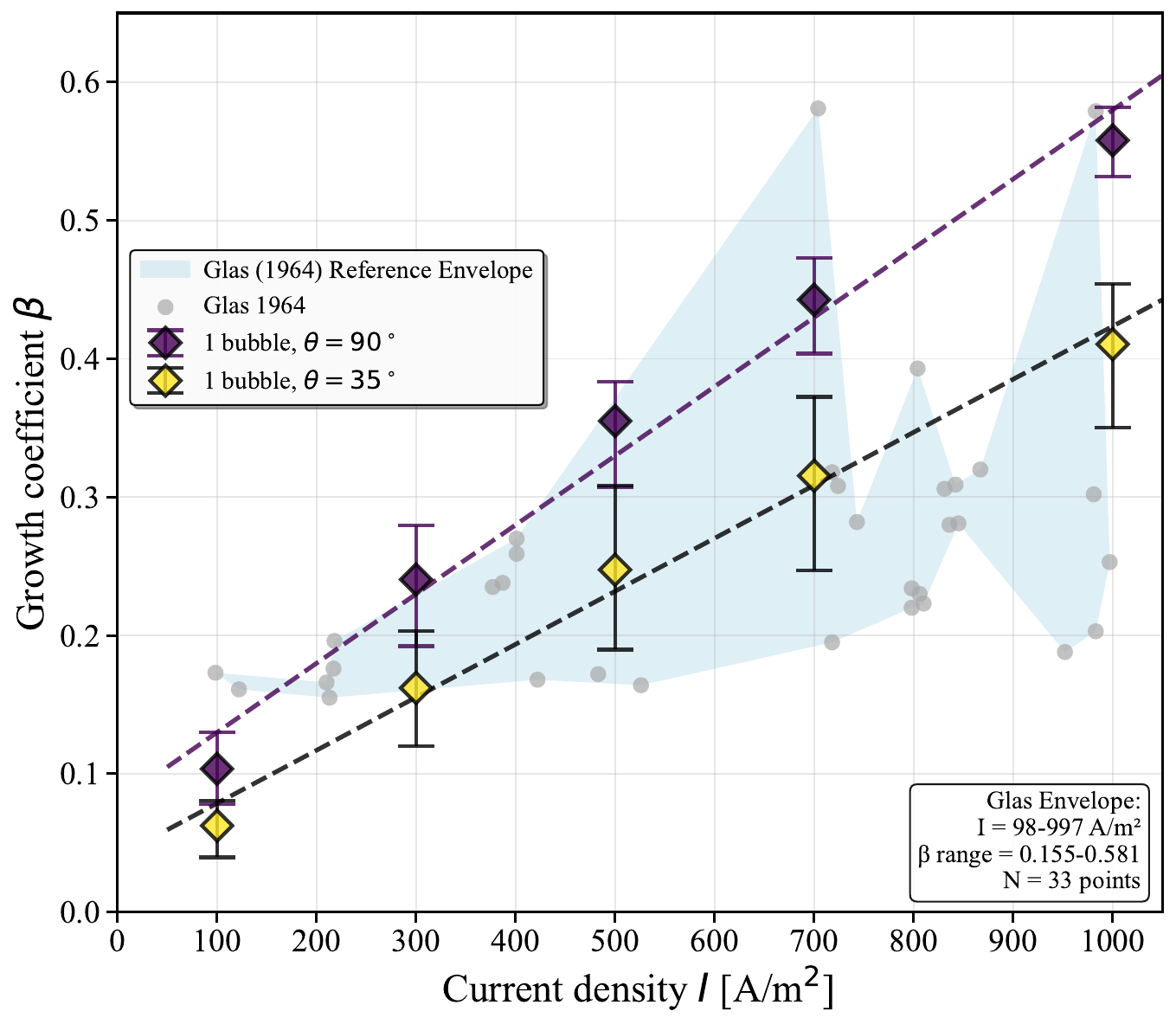}
    \caption{Comparison of growth rate of a single bubble between experimental and numerical results.}
    \label{fig: glas_compare}
\end{figure}

From the analytical solution \(R=2\beta t^{1/2} \), 
the growth rate \(\beta\) can be determined. The fitting method used by \cite{Glas1964} involves plotting the bubble radius \(R\) versus \(t^{1/2}\), from which $\beta$ is computed directly from the slope.
Using the same historical fitting approach, the numerical results showing the influence of current density on the growth rate are presented in \cref{fig: glas_compare}. Because the bubble growth exponent does not perfectly match $1/2$ during the early growth stage (as discussed in \cref{fig: ele_log}), we performed three local fittings at different times for each case, with the resulting error bar indicating the variability among these estimates.
The results clearly show that the growth rate increases with increasing current density. This trend is expected, as the current density directly dictates the hydrogen flux at the electrode surface, which drives bubble growth through higher concentrations of dissolved hydrogen near the electrode.

Then, we find that a bubble with a contact angle of $\theta = 90^\circ$ grows faster than one with a contact angle of $\theta = 35^\circ$. This difference can be explained by considering the influence of the contact angle on the bubble shape and placement. Mass transfer through diffusion occurs only through the interface between the gas and liquid phases. 
In the $\theta = 90^\circ$ scenario, the bubble elongates more along the electrode wall, 
thereby increasing its exposure to the high-concentration region close to the electrode. 
As a result, the effective diffusive mass transfer to the bubble with \(\theta =90 \degree\) is enhanced. 

\begin{figure}[htbp]
    \centering
    \includegraphics[width=0.9\linewidth]{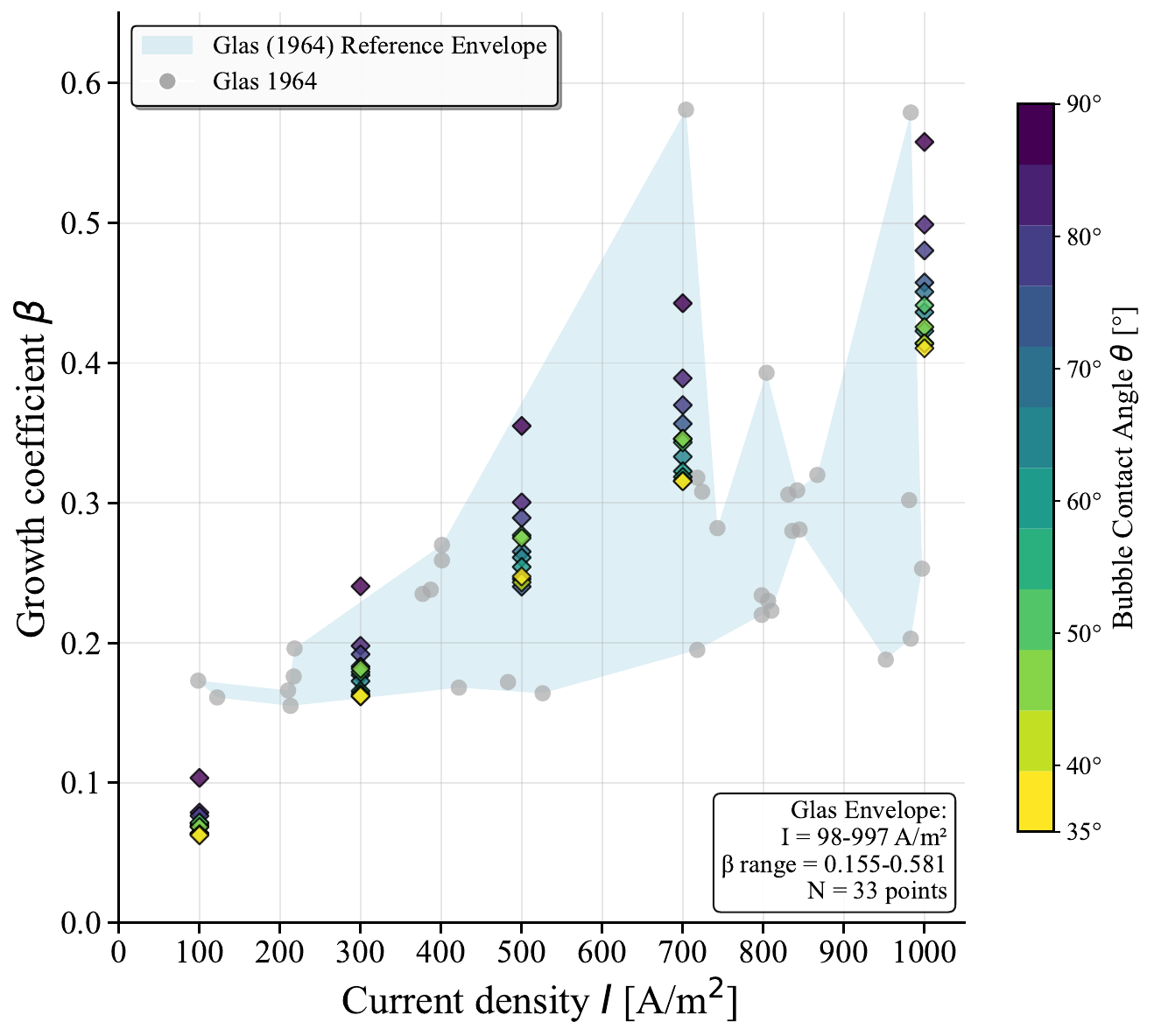}
    \caption{Comparison of growth rate of a single bubble between experimental and numerical results for various contact angles.}
    \label{fig: glas_compare_angle}
\end{figure}

Furthermore, \cref{fig: glas_compare_angle} presents the growth rate trend deduced from several different contact angles. The simulations cover the range from $35^{\circ} \le \theta \le 90^{\circ}$ for the same initial nucleation diameter. This range is relevant as most electrode surfaces are designed to be hydrophilic to facilitate the removal of attached bubbles. The results indicate that the bubble-growth coefficient generally increases with the contact angle, until the upper limit of our simulations of $\theta=90^{\circ}$. However, as the contact angle approaches the lower limit, the bubble-growth coefficient is no longer dominated by the contact angle. In some instances, such as at a current density of $I=500\,\mathrm{A/m^2}$, this trend is even reversed. While we have no explanation for this reversal, we attribute the weakening influence of the contact angle near the lower limit to a 
geometrical effect. Indeed, as the contact angle decreases, the bubble geometry changes only a little while it approaches a perfect sphere, and the influence of the contact angle on growth becomes less significant. Consequently, simulation results for contact angles smaller than $\theta=35^{\circ}$ are not shown here.

A significant discrepancy in the effect of current density is also noted when comparing the simulated and experimentally measured bubble-growth coefficients $\beta$, as shown in \cref{fig: glas_compare}. For the $\theta=35^\circ$ case, the numerical model underestimates $\beta$ at low current densities. The deviation is more complex for $\theta=90^\circ$, where $\beta$ is underestimated at low current densities but is conversely overestimated at higher current densities. This discrepancy is observed not only in the present simulation but also in the results of \cite{Gennari2022}.

The underestimation is likely due to the difference between the numerical setting and the experimental conditions. \cite{Glas1964} established a nonlinear relationship between the growth coefficient $\beta$ and the current density $I$, as the driving force (local supersaturation) inherently depends on the current density. This model $\beta=I^a, a<1$ effectively describes the experimental minor change of $\beta$ observed for $(100 < I < 500\,\mathrm{A/m^2})$. In contrast, our numerical work assumes that the rate of $\mathrm{H_2}$ production is directly proportional to the electrical current. This simplified setting cannot perfectly replicate the real situation, where bubble production at any site is intermittent in practice. Furthermore, the simulation uses a typical nucleation time of $0.02\mathrm{s}$. This time point determines the concentration layer of dissolved hydrogen when bubble growth begins. However, the nucleation time in experiments can vary between $0.02 \mathrm{s}$ to $0.1\mathrm{s}$ \citep{Glas1964}. The difference in nucleation time at low current densities results in a weaker hydrogen concentration boundary layer than the experimental boundary layer, and therefore a smaller growth coefficient.

To explain the overestimation observed in the simulations, two major differences between the simulated and experimental conditions should be noted.
First, the contact angle is found to change from \(70\degree\) to \(20 \degree\) in the experiment. Since our simulations utilize a constant contact angle, a perfect match across all experimental conditions is not expected. However, it is noteworthy that the discrepancy is minimal when comparing the simulation results within the experimentally observed range of \(35\degree\leq\theta\leq70\degree\).

Second, obtaining a truly isolated, single active nucleation site on an electrode surface is challenging in both modern experiments and earlier studies \citep{Glas1964}. This suggests that experimental measurements of presumed single-bubble growth may, in fact, be influenced by neighboring bubbles, such as a bubble carpet or multiple nucleation sites. When multiple nucleation sites are active (a condition dependent on the electrode material and its surface properties), the dissolved hydrogen in the liquid is consumed simultaneously by several bubbles. Consequently, the individual growth rates are reduced compared with the idealized case of a single nucleation site, which benefits from the full hydrogen flux from the electrode, contributing to the growth of one bubble alone.

\subsection{Multiple bubbles growth}\label{sec: multi_growth}

There is a notable decrease in the growth rate and exponent of hydrogen bubbles when multiple nucleation sites are active \citep{Glas1964}. 
This Section numerically investigates the mutual bubble interactions for multiple nucleation sites. 
Instead of a single bubble, we model two bubbles and four bubbles growing at the same electrode. Since the influence of the contact angle on bubble growth has already been discussed in \cref{sec: single_growth}, the results presented in this section focus on bubbles growing at a contact angle of \(\theta=90\degree\).

\begin{figure}[htbp]
    \centering
    \includegraphics[width=0.9\linewidth]{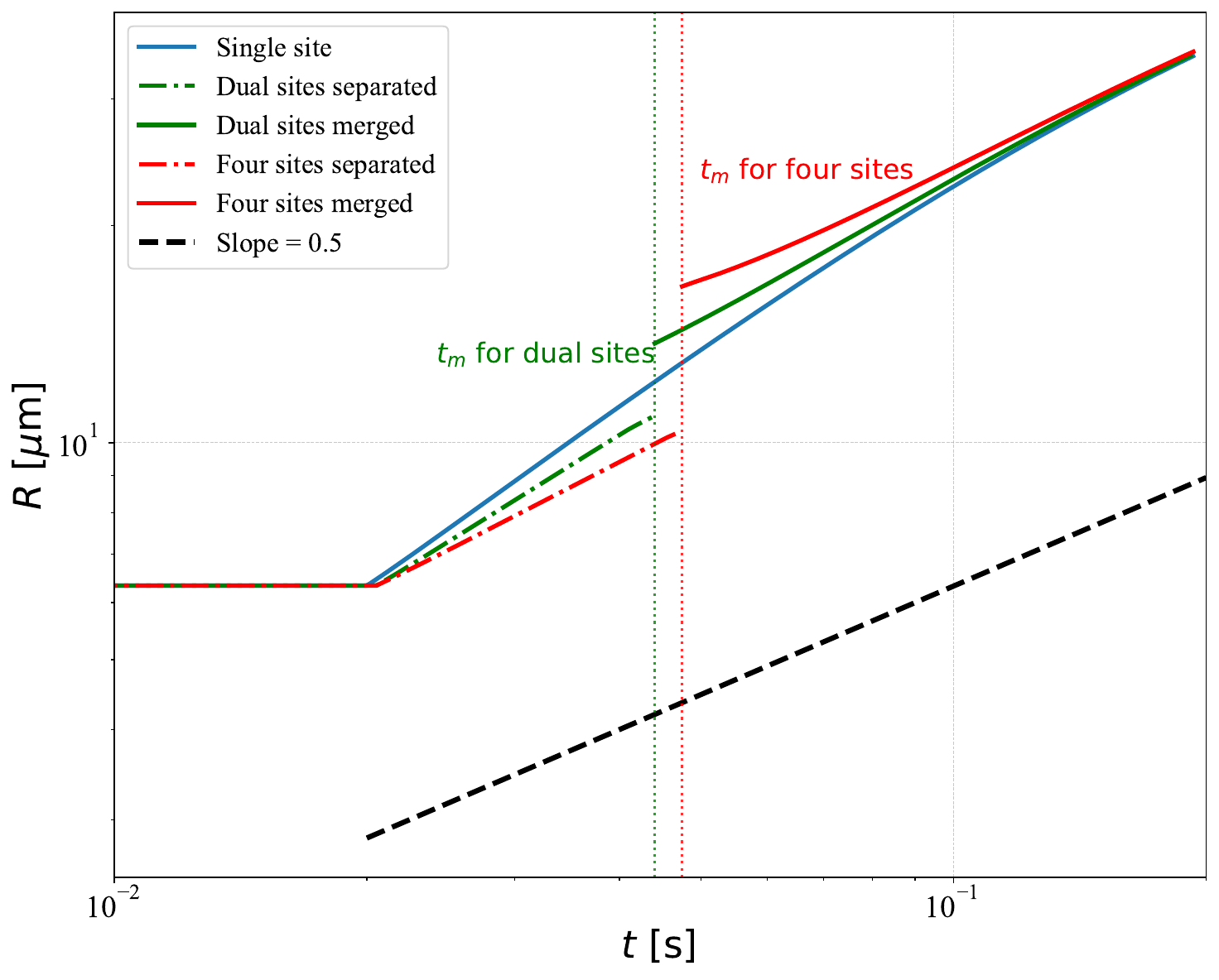}
    \caption{Growth of multi-bubbles at the contact angle of \(\theta=90\degree\). The current density is \(\mathrm{1000 A/m^2}\) and the slope represents the growth exponent.}
    \label{fig: multi}
\end{figure}

The growth of multiple nucleation sites is mutually suppressed in the stage before bubble coalescence from the growth slope \cref{fig: multi}. 
During this stage, hydrogen is consumed by all the bubbles simultaneously, leading to a reduced growth rate and exponent. The larger the number of active nucleation sites, the slower the growth.
However, an unexpected behavior of the bubbles' growth seems puzzling. 
Intuitively, once the bubbles merge, they should approximate the growth dynamics of a single bubble case.
Contrary to this assumption, even after bubble coalescence(\(t>t_m\)), differences in the growth persist for a period. Then the growth curves of all cases converge, as shown in \cref{fig: multi}. 

A possible explanation lies in the relative position of bubbles within the hydrogen concentration layer. As shown in \cref{fig: multi},
the trend remains consistent: the more active nucleation sites, the slower the growth, regardless of the stage (before or after bubble coalescence).
The key reason for the different growth behavior after coalescence is the bubble radius. 
Observations indicate that the radius of the merged bubble in the four-nucleation-site case is larger than in the dual-site case, 
while the bubble in the single-site case has the smallest radius. Despite this, 
the hydrogen concentration layer remains nearly identical across different nucleation-site cases at the same time. 
This is because the hydrogen flux at the boundary is sufficiently high, making the amount of hydrogen consumed by bubble growth negligible.
Then, the crucial factor is immersion depth: bigger bubbles immersed less deeply into the hydrogen concentration layer, leading to a lower mass transfer rate, which means a smaller growth exponent, as illustrated in \cref{fig: multi}. 

\begin{figure}[htbp]
    \centering
    \begin{subfigure}[b]{0.48\linewidth}
        \centering
        \includegraphics[width=\linewidth, trim={1680 0 1680 0}, clip]{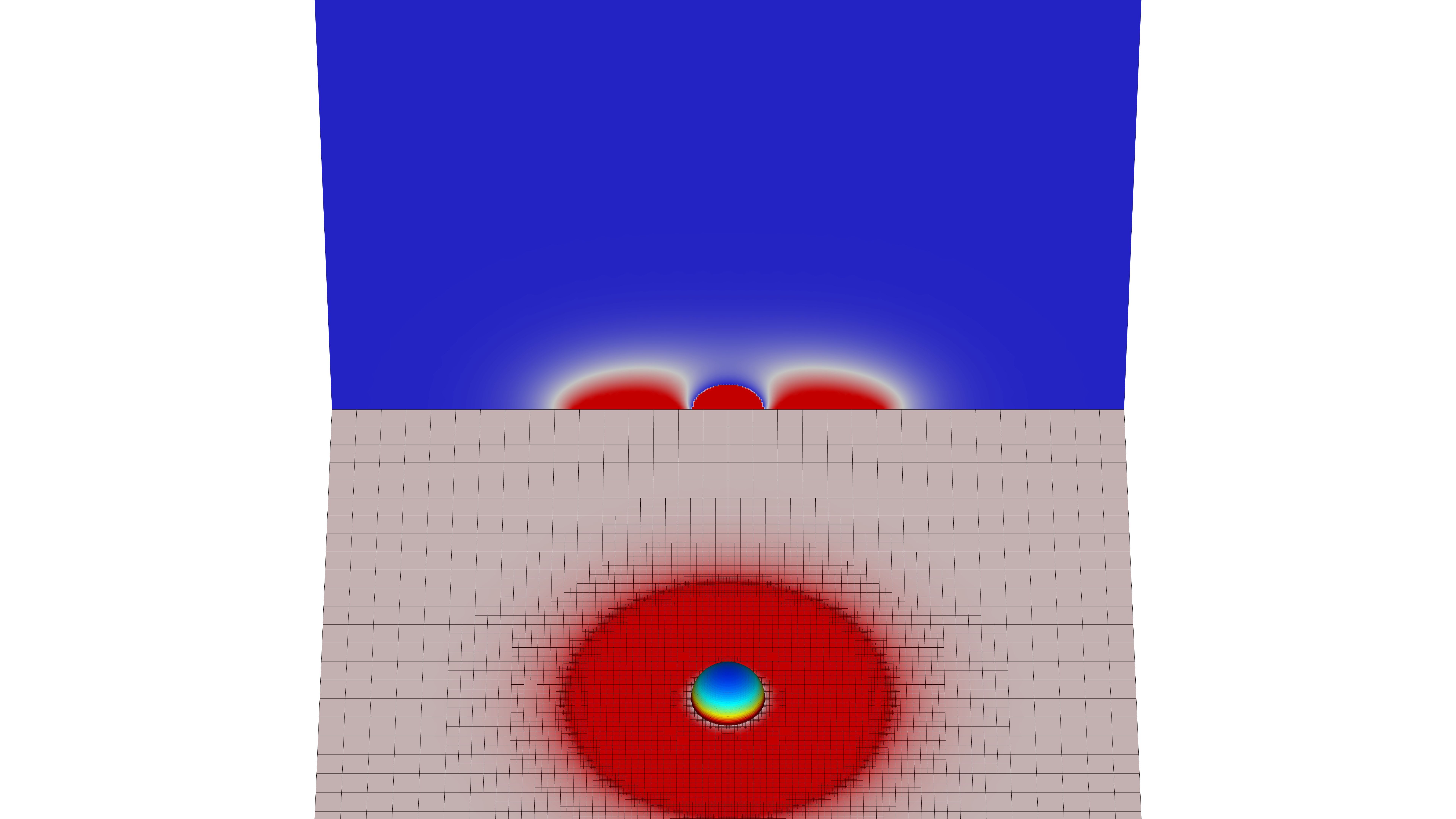}
        \caption{Single nucleation site at merge time \(t_m\).}
        \label{fig: 1_280}
    \end{subfigure}
    \hfill
    \begin{subfigure}[b]{0.48\linewidth}
        \centering
        \includegraphics[width=\linewidth, trim={1680 0 1680 0}, clip]{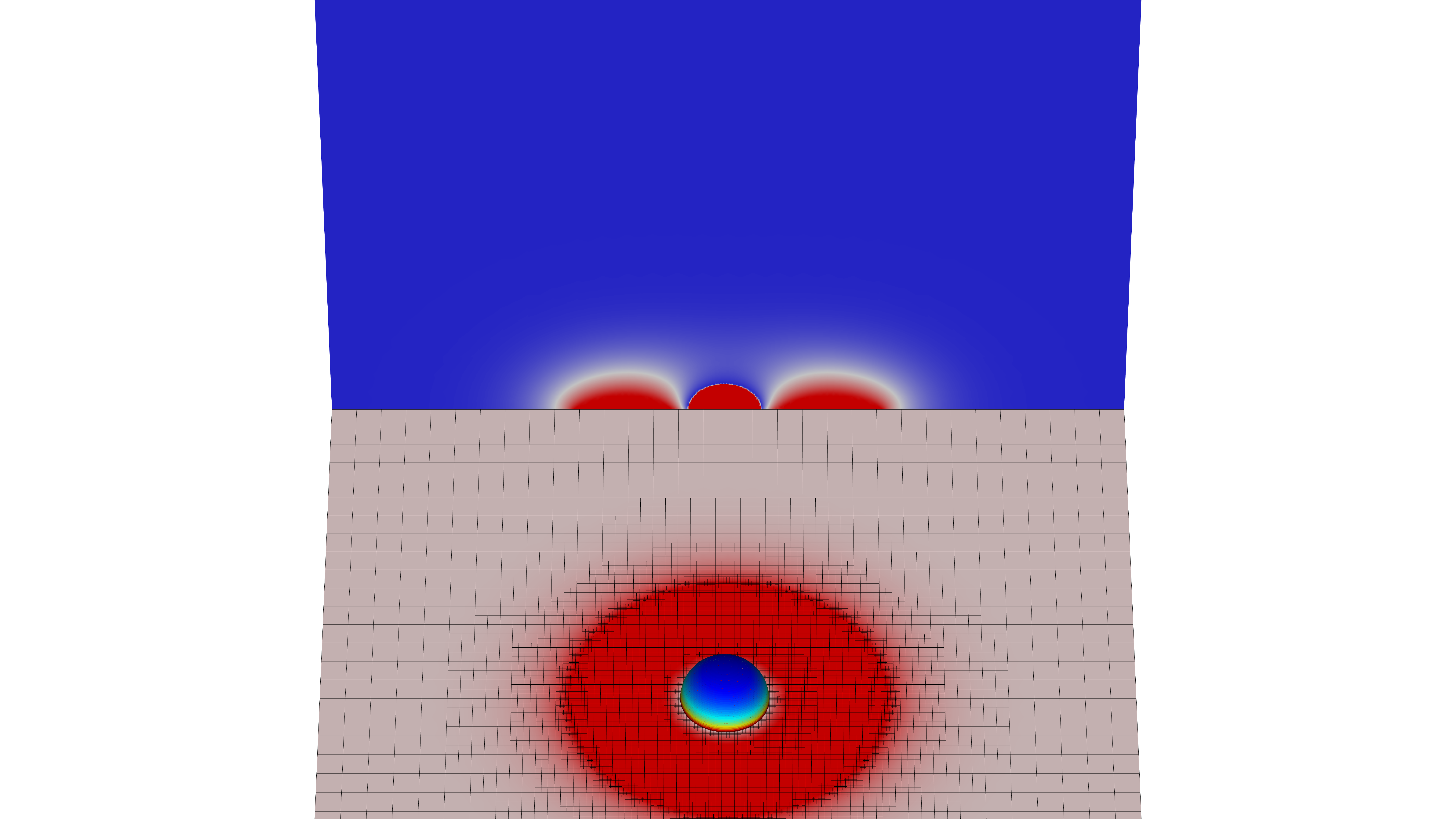}
        \caption{Four nucleation sites at merge time \(t_m\).}
        \label{fig: 4_280}
    \end{subfigure}
    \caption{Frames show the comparison between the single bubble and four bubbles at the contact angle of \(\theta=90\degree\). The current density is 1000 A/m$^2$. The view direction is the same as in \cref{fig: sketch_3D}. The hydrogen contour on the intersecting plane (yz) passing through the middle of the bubble is projected to the back plane. The bubble interface is colored according to the local mass-transfer rate.}
    \label{fig: 1b_4b_compare}
\end{figure}

Supporting evidence emerges from direct comparisons: at the same time $t_m$, the four-nucleation-site configuration produces marginally larger bubbles than the single-site case, as shown in \cref{fig: 1b_4b_compare}.

However, the constant hydrogen flux induces a countervailing effect. As illustrated by the snapshots in \cref{fig: 1b_4b_compare}, where the bubble surface is colored by the local mass transfer rate, a higher overall mass transfer rate slows the increase of the hydrogen concentration boundary layer for the four-nucleation-site case. This effect enables shallower bubble immersion after coalescence. This feedback mechanism drives a gradual decrease in the growth exponent, ultimately causing convergence across different nucleation-site configurations. As shown in \cref{fig: multi}, the growth exponent gradually approaches a slope of $1/2$.

\begin{figure}[htbp]
    \centering
    \includegraphics[width=0.9\linewidth]{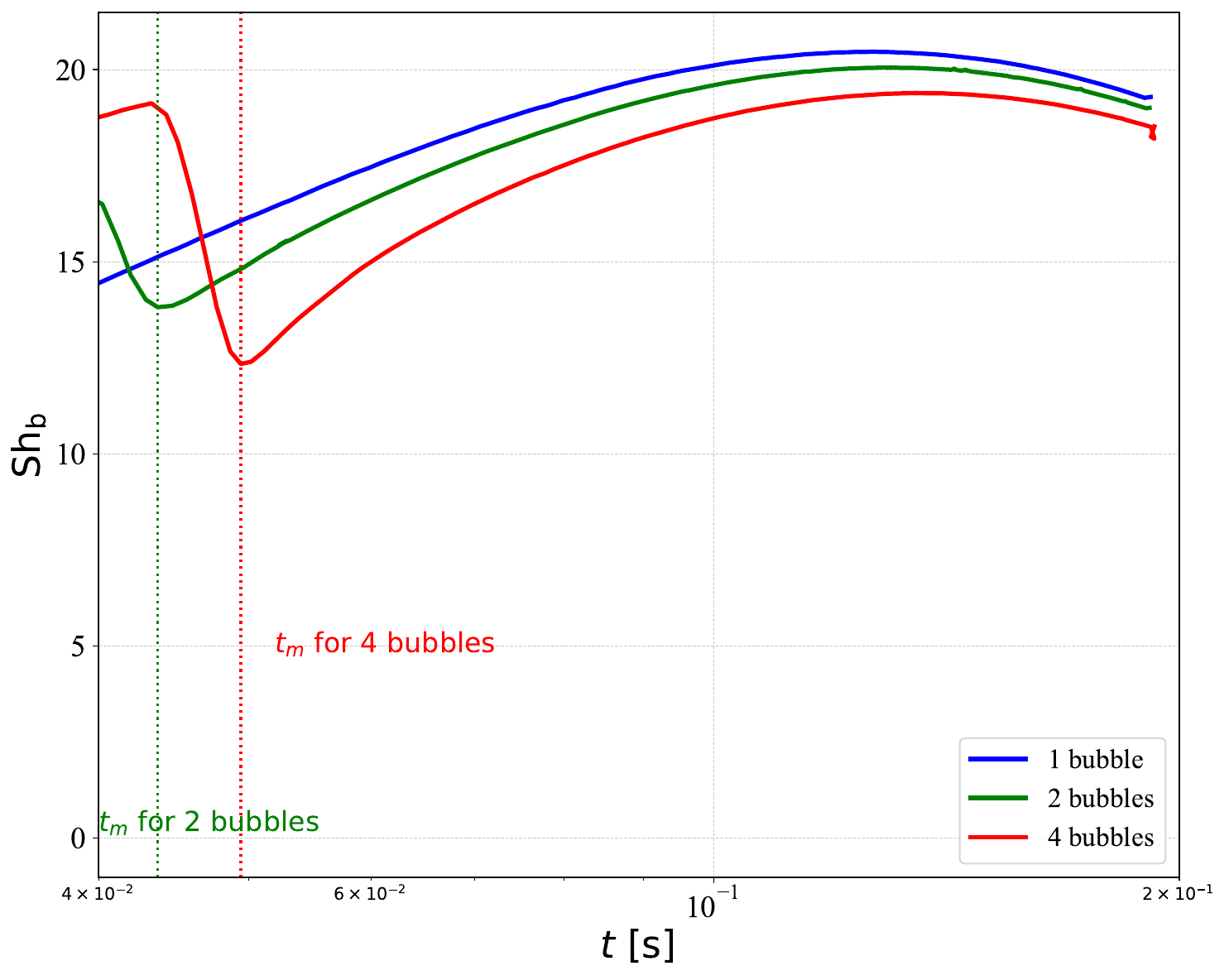}
    \caption{The Sherwood number for simulations with different numbers of nucleation sites. The current density is $1000 \mathrm{A/m^2}$.}
    \label{fig: multi_sh}
\end{figure}

To quantitatively assess mass transfer, \cref{fig: multi_sh} allows to analyze the time-dependent bubble Sherwood number defined in \cref{eq: sh_b}, 
which quantifies mass transport efficiency to the bubble.
It should be noticed that multiple nucleation sites exhibit elevated \(\mathrm{Sh_b}\) initially. A decrease occurs after bubble coalescence(\(t_m\)), which causes the \(\mathrm{Sh_b}\) to be smaller than the one for the single nucleation site.
The plot of \(\mathrm{Sh_b}\) cross-verified the explanation of the difference in growth rate above.
First, the mutual suppression. While total mass transport increases with nucleation sites, individual bubble transport efficiency is inhibited through competitive \(\mathrm{H_2}\) consumption.
Second, the immersion depth: lower after-merge \(\mathrm{Sh_b}\) directly correlates with diminished mass transfer rates for multiple nucleation sites, consistent with the immersion depth arguments.

\begin{figure}[htbp]
    \centering
    \includegraphics[width=0.9\linewidth]{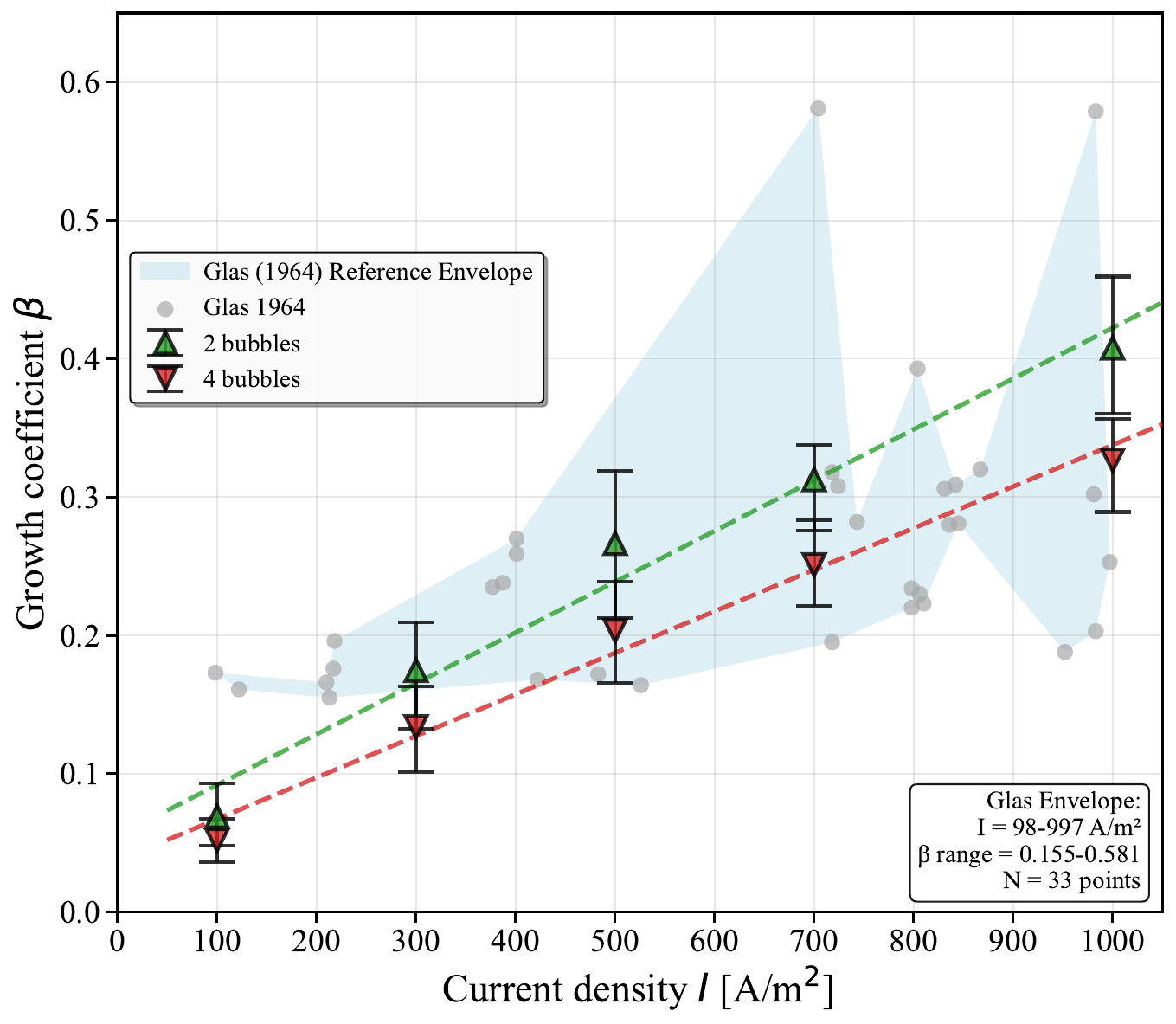}
    \caption{Comparison of growth rate for multiple nucleation sites with the contact angle of \(\theta=90 \degree\).}
    \label{fig: multi_rate}
\end{figure}

As shown in the comparison between experimental and simulation results in \cref{fig: multi_rate}, the growth rates obtained from the multiple-bubble cases exhibit better agreement with experimental data than those from the single-bubble case (\cref{fig: glas_compare}). This observation supports the earlier conjecture regarding the influence of neighboring bubbles. Although the reported experimental growth rate used for comparison was acquired for a single bubble attached to the electrode, likely, small bubbles forming underneath were not captured during measurement. Consequently, the growth rates predicted by the multiple-bubble simulations may more accurately represent the actual experimental conditions.

Furthermore, the experimental data itself exhibits a considerable standard deviation, and the growth coefficient $\beta$ was obtained through a fitting process without associated error bars. This experimental uncertainty presents an alternative explanation for the observed differences between simulation and experimental measurement.
\begin{figure}[htbp]
    \centering
    \includegraphics[width=0.9\linewidth]{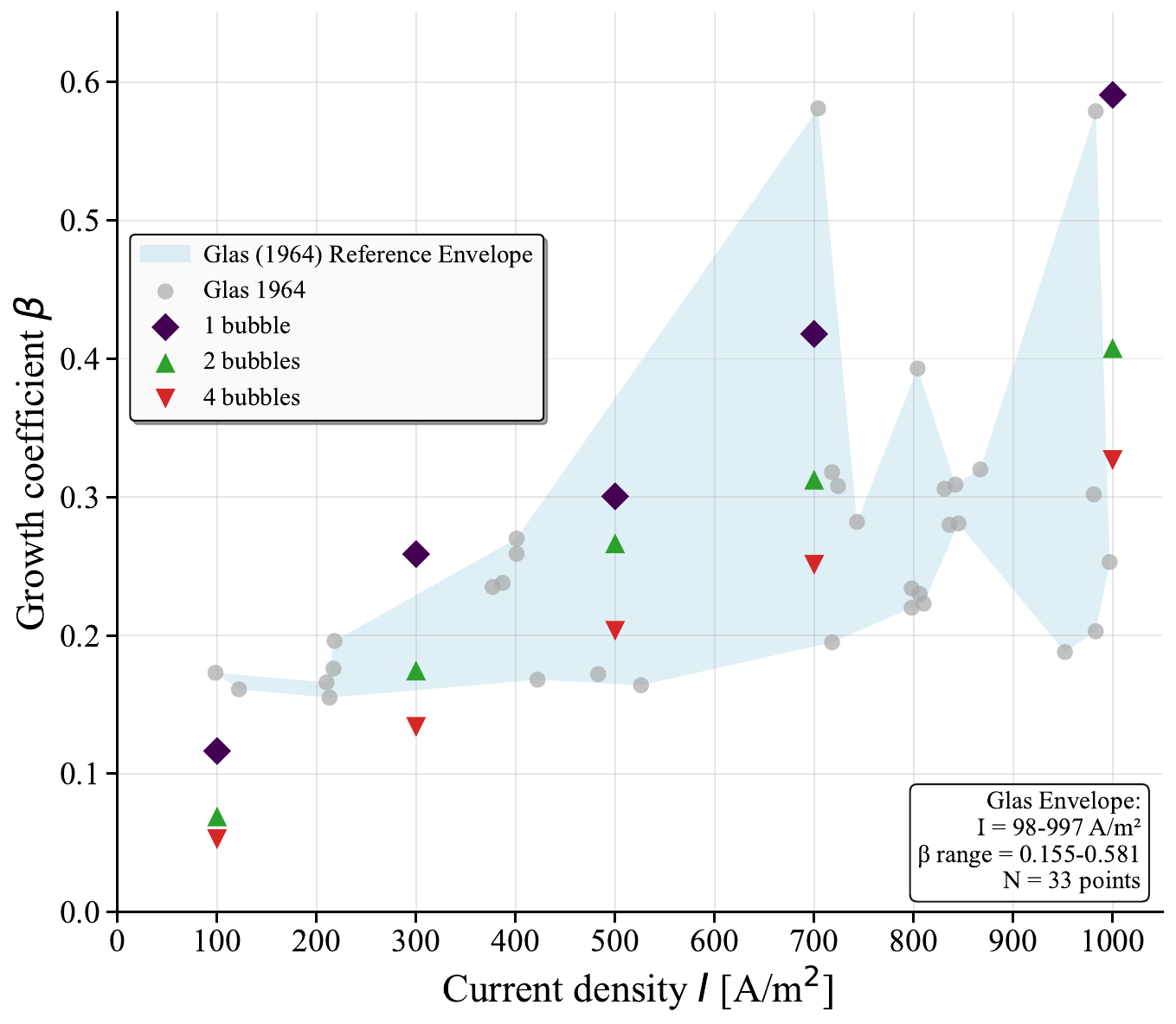}
    \caption{Comparison of growth rate for single site and multiple nucleation sites with the contact angle of \(\theta=90 \degree\).}
    \label{fig: multi_compare}
\end{figure}

To comprehensively assess the influence of nucleation sites, \cref{fig: multi_compare} compares the average bubble growth coefficients $\beta$ for one, two, and four bubbles. In addition to the expected suppression of the growth coefficient, the results reveal that the relative impact of multiple nucleation sites becomes more significant at higher current densities. This observation may help explain the earlier discrepancies between simulations and experiments: the higher current density amplifies the effect of other parameters, such as contact angle and the density of nucleation sites. Notably, the overestimation of the bubble-growth coefficient is remarkable at the current density of $I=1000\,\mathrm{A/m^2}$, whereas at lower current densities $(300 < I < 800\,\mathrm{A/m^2})$, the numerical predictions are comparable to the experimental measurements.
\subsection{Bubble detachment}
The removal of attached bubbles is crucial for enhancing the efficiency of the electrolysis process \citep{Angulo2020}. 
In the absence of contact angle hysteresis, the bubble tends to spread on the electrode surface with constant speed. 
The force balance determines the relation between the volume-equivalent detachment radius (Fritz radius) and the value of the contact angle $\theta$ giving\citep{Stephan1979},
\begin{align}
    R_{det}=0.6\, \theta \sqrt{\frac{\sigma} {(\rho_c-\rho_d)g}}.
    \label{eq: Fritz_stephane}
\end{align}
%The general contact angle for the bubble at the electrode would be smaller than \(\pi/2\). 
This analytical solution is derived by balancing the buoyancy force for a perfect sphere-shaped bubble $F_b=(4/3)\pi R^3(\rho_c-\rho_d)g$ with the capillary force for a bubble attached 
to the electrode. %The contact angle controls the magnitude of surface tension effects and varies in connection with contact line dynamics, which are still, in many circumstances, an object of investigation. Molecular dynamics simulations with either Lennard-Jones intermolecular potentials or more realistic water molecule models give very different results, see \cite{lacis2022a}. Therefore, the theoretical value is adopted as a reference to estimate the order of magnitude rather than for precise comparison in the present study.

To begin with, a simulation of single-bubble detachment is performed.
Initially, the bubble remains nearly spherical due to its small size, which results in a negligible buoyancy force. However, as the bubble radius increases, 
deformation becomes evident. Eventually, buoyancy overcomes surface tension, leading to the detachment from the electrode surface.
\begin{figure}[htbp]
    \includegraphics[width=0.9\linewidth]{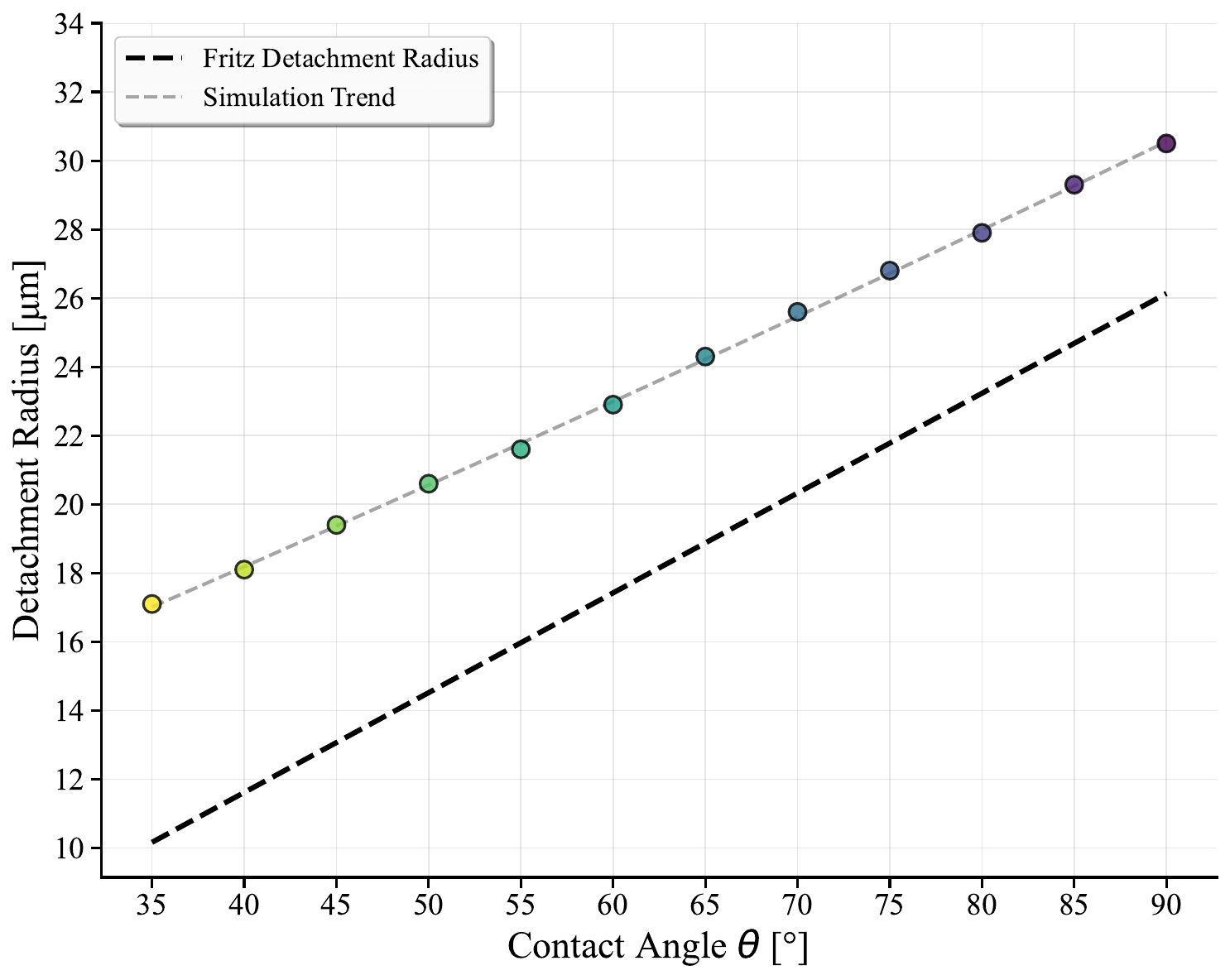}
    \caption{Bubble detachment radius versus contact angle for a single bubble(with the current density of \(I=\mathrm{1000 A/m^2}\)).}
    \label{fig: Fritz_validate}
\end{figure}
\begin{figure*}[htbp]
    \includegraphics[width=0.9\linewidth]{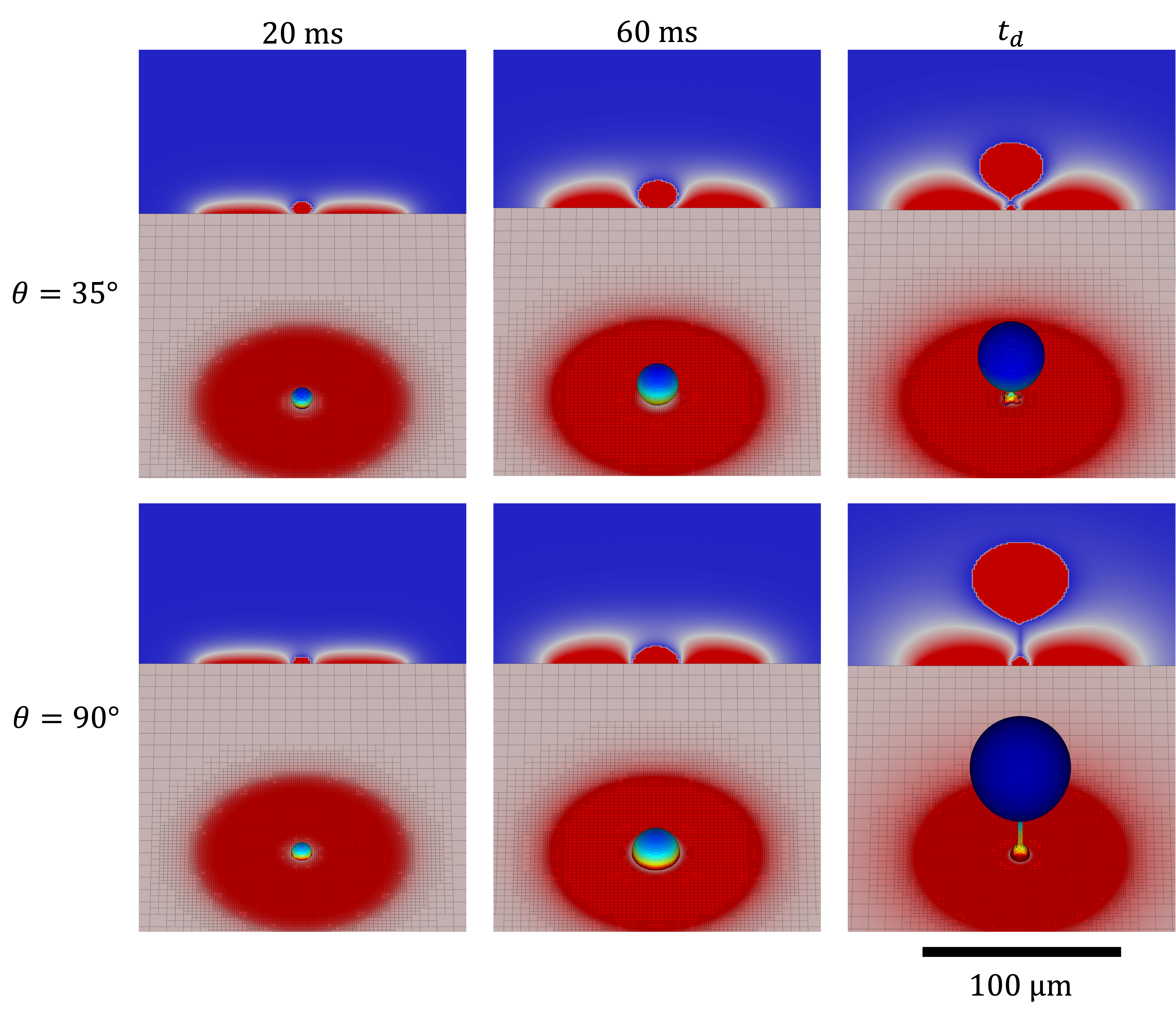}
    \caption{Frames show the detachment evolution for a single bubble at different contact angles(with the current density of \(I=\mathrm{1000 A/m^2}\)). The corresponding detachment times ($t_d$) are $104.2 \mathrm{ms}$ for $\theta=35^\circ$, and $243.2 \mathrm{ms}$ for $\theta=90^\circ$. The hydrogen contour on the intersecting plane (yz) passing through the middle of the bubble is projected to the back plane. The bubble interface is colored according to the local mass-transfer rate.}
    \label{fig: detach_35_90}
\end{figure*}

As previously mentioned, \cref{eq: Fritz_stephane} establishes a linear relationship between \(R_d\) and \(\theta\). To further investigate this dependency, a series of cases with varying contact angle from \(\theta=35\degree\) to \(\theta=90\degree\) are simulated, see \cref{fig: Fritz_validate}. As expected, the numerical results also exhibit a linear relationship. The Fritz radius is computed using the same reduced surface tension. However, the simulation predicts larger detachment radii compared to the theoretical value. The pinning effect of contact angle hysteresis can explain this deviation. For detachment to occur, the contact line must recede (shrink inwards). However, contact angle hysteresis pins the contact line. The bubble cannot simply shrink at its base; instead, its neck must stretch and thin until the surface tension force can no longer hold it. Since the theoretical Fritz radius is derived from a simplified force balance that neglects contact angle hysteresis, the theoretical value is expected to be smaller than the detachment radius predicted by the simulation.

To demonstrate the effect of contact angle on detachment radius, we take the most contrasting cases, $\theta=90^{\circ}$ (hydrophobic) and $\theta=35^{\circ}$ (hydrophilic), as an example. As shown in \cref{fig: detach_35_90}, the bubble with the smaller contact angle ($\theta=35^{\circ}$) detaches significantly faster and at a much smaller radius.

\begin{figure}[htbp]
    \includegraphics[width=0.9\linewidth]{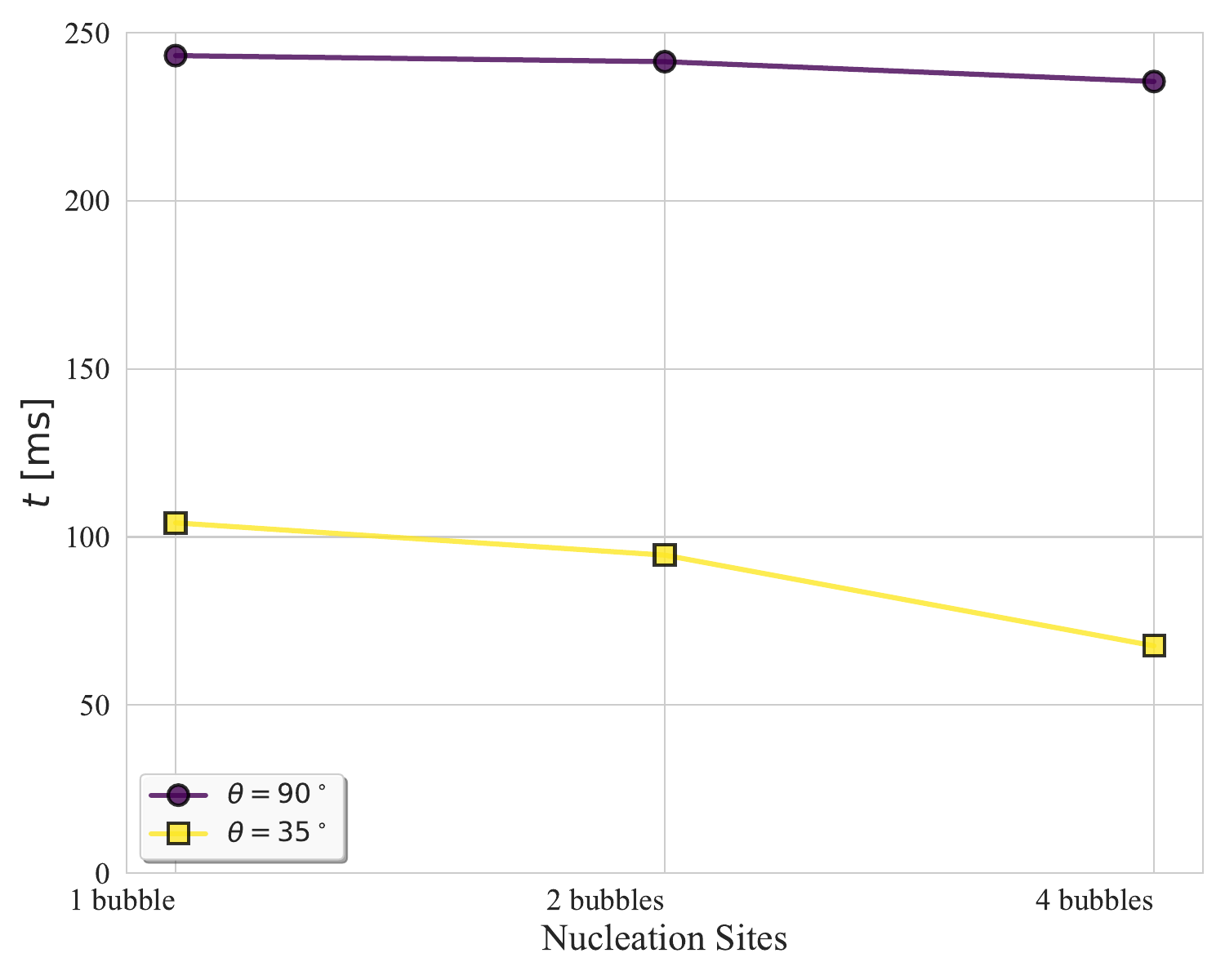}
    \caption{Bubble detachment times for different nucleation sites and contact angles (with the current density of \(I=\mathrm{1000 A/m^2}\)).}
    \label{fig: multi_sites_time}
\end{figure}

\begin{figure}[htbp]
    \includegraphics[width=0.9\linewidth]{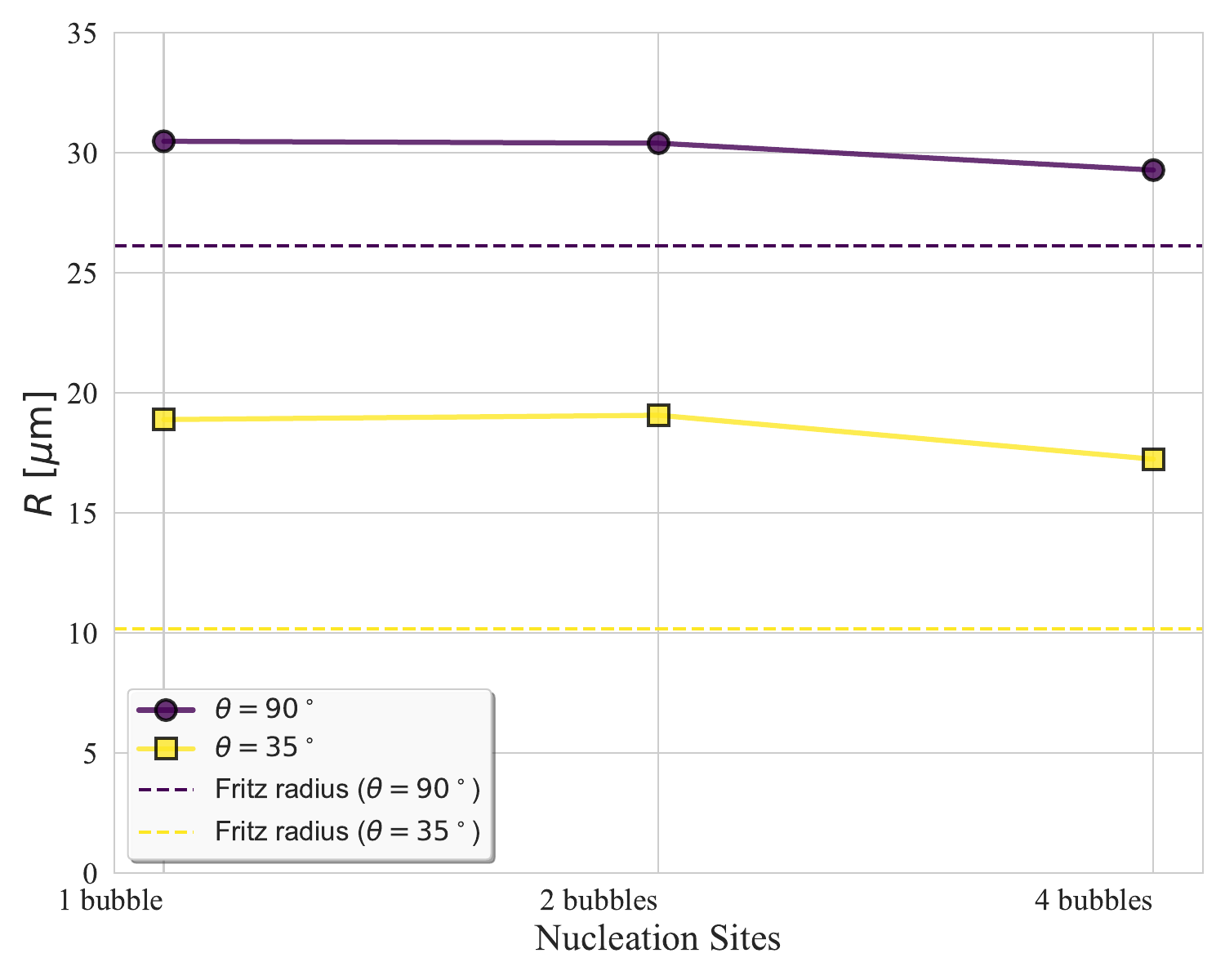}
    \caption{Bubble detachment radius for different nucleation sites and contact angles (with the current density of \(I=\mathrm{1000 A/m^2}\)).}
    \label{fig: multi_sites_radius}
\end{figure}

Additionally, it has been shown that mutual interactions between bubbles influence the detachment process. 
\citet{Bashkatov2024} observed that bubble coalescence in the presence of dual bubbles can lead to significantly 
earlier detachment and a smaller detachment radius compared to cases where only buoyancy effects are considered.
%To verify this effect, we conduct a series of simulations involving multiple nucleation sites. 
As illustrated in \cref{fig: multi_sites_time}, the presence of multiple bubbles indeed accelerates the detachment. Consequently, the detachment radius is reduced, as demonstrated in \cref{fig: multi_sites_radius}.

\begin{figure*}[htbp]
    \includegraphics[width=0.9\linewidth]{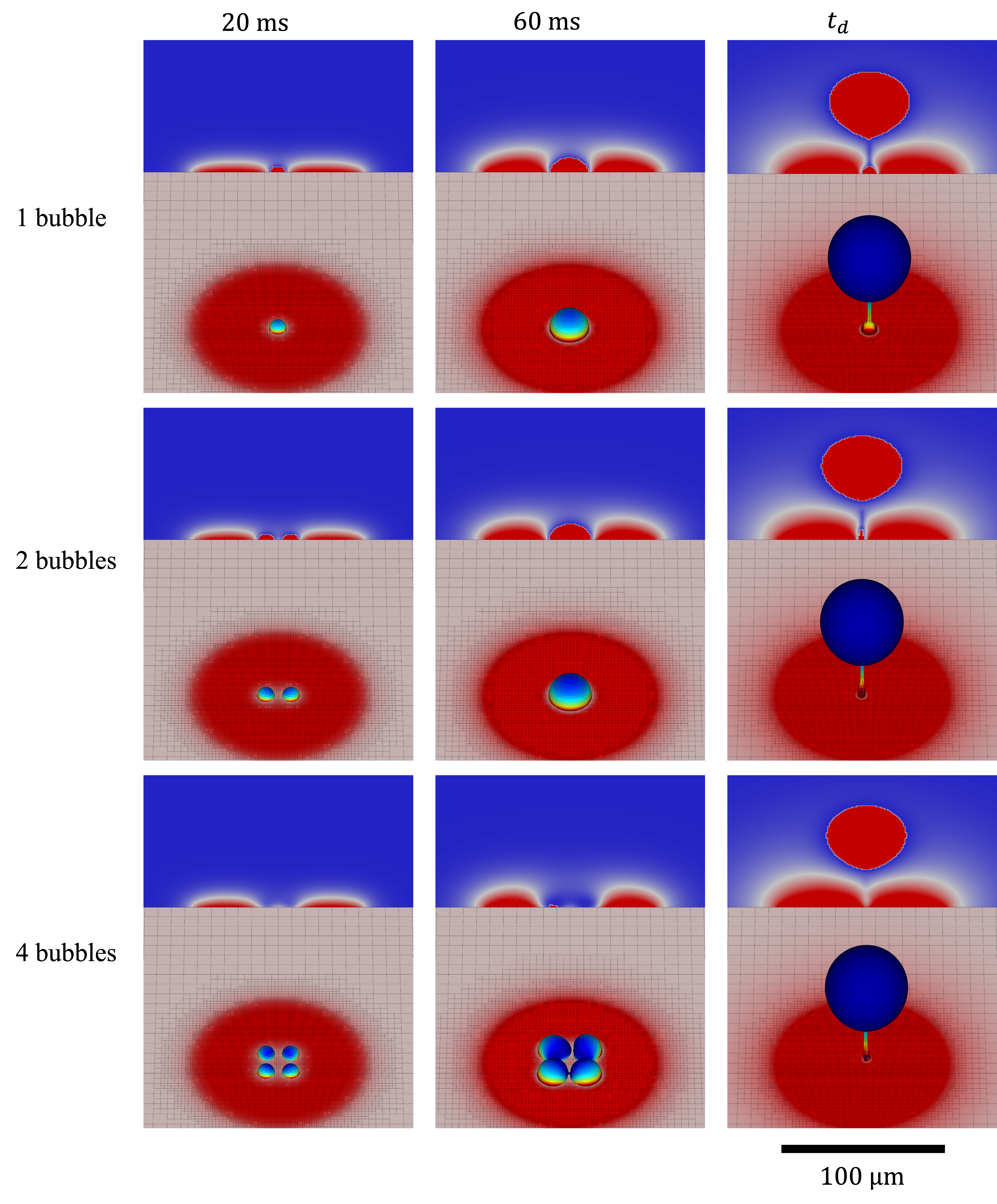}
    \caption{Frames show the detachment evolution for multi-bubbles at the contact angle of \(\theta=90\degree\) (with the current density of \(I=\mathrm{1000 A/m^2}\)). The corresponding detachment times ($t_d$) are $243.2 \mathrm{ms}$ for 1 bubble, $241.4 \mathrm{ms}$ for the 2-bubbles case, and $235.5 \mathrm{ms}$ for the 4-bubbles case. The hydrogen contour on the intersecting plane (yz) passing through the middle of the bubble is projected to the back plane. The bubble interface is colored according to the local mass-transfer rate.}
    \label{fig: detach_3D_90}
\end{figure*}

\begin{figure*}[htbp]
    \includegraphics[width=0.9\linewidth]{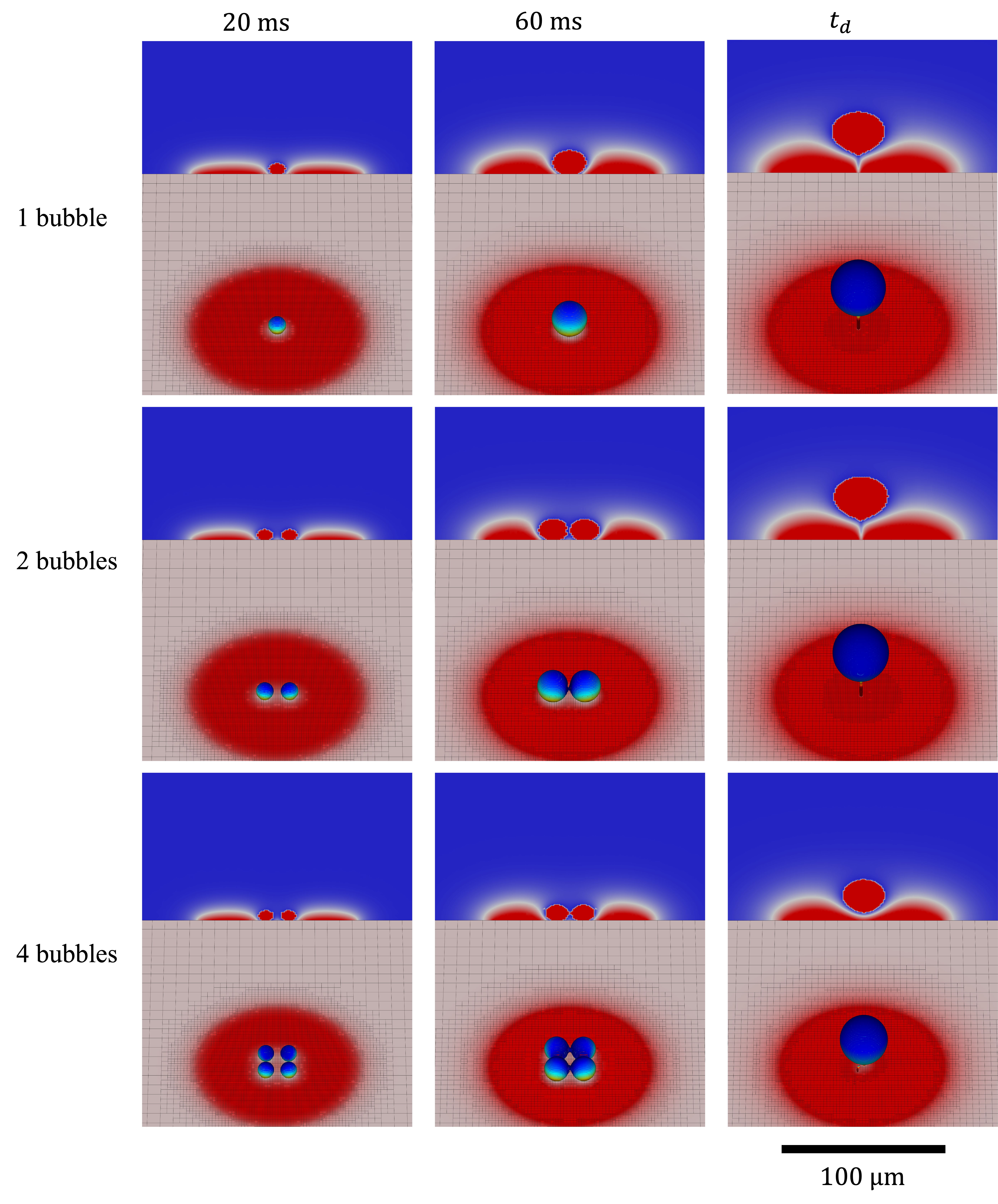}
    \caption{Frames show the detachment evolution for multi-bubbles at the contact angle of \(\theta=35\degree\) (with the current density of \(I=\mathrm{1000 A/m^2}\)). The corresponding detachment times ($t_d$) are $104.2 \mathrm{ms}$ for 1 bubble, $94.6 \mathrm{ms}$ for the 2-bubbles case, and $67.6\mathrm{ms}$ for the 4-bubbles case. The hydrogen contour on the intersecting plane (yz) passing through the middle of the bubble is projected to the back plane. The bubble interface is colored according to the local mass-transfer rate.}
    \label{fig: detach_3D_35}
\end{figure*}

For a direct visual comparison, representative frames from simulations with multiple bubbles at contact angles of $\theta = 90^\circ$ and $\theta = 35^\circ$ are provided in \cref{fig: detach_3D_90,fig: detach_3D_35}.
A plausible mechanism behind this acceleration is as follows. The coalescence-induced shape perturbations generate an initial momentum favoring premature detachment
from the electrode surface. It is observed in Bashkatov's experiment that the coalescence of bubbles results in an initial jump-off of the merged bubble.  
This can be attributed to the released surface energy during coalescence \citep{Zhang2024}. The released energy is partly dissipated by the bubble oscillations, working against viscous drag. 
When in the proximity to the surface, the remaining energy is converted to kinetic energy, driving the resultant (merged) bubble to jump off the electrode \citep{lv2021}.
This kinetic energy can cause bubble departure at smaller radii than in the purely buoyancy-driven scenario, as observed in our simulation results.
However, the capillary waves due to energy dissipation are not observed in our simulations because of the mesh resolution constraint.
Further numerical investigations are needed to distinguish the respective roles of buoyancy and interfacial effects in this process.

\section{Conclusion}\label{sec:conclusions}

This study investigates the mechanisms of bubble growth, coalescence, 
and detachment at horizontal electrodes through three-dimensional direct numerical simulations. 
A convergence study is conducted to verify numerical accuracy, with results compared against experimental data. 
Key parametric analyses include the effects of bubble contact angle and multi-bubble interactions. 
The following conclusions are drawn:

\noindent\textbf{Bubble growth dynamics}  

The bubble contact angle significantly influences growth behavior by dictating the geometric position of the bubble within the evolving hydrogen concentration profile. This geometric position directly modulates the local mass-transfer rates.

For multi-bubble systems, mutual suppression of mass transport occurs due to the simultaneous hydrogen consumption by neighboring bubbles, a phenomenon quantified by reductions in the Sherwood number ($\mathrm{Sh}_b$), as shown in \cref{fig: multi_sh}. Specifically, increasing the number of nucleation sites leads to a smaller individual bubble Sherwood number. While some numerical overprediction and underprediction of growth rates exist compared to experimental observations (\cref{fig: multi_rate}), the simulations adhere closely to the scaling law of Scriven's solution ($R \propto t^{1/2}$).

\noindent\textbf{Bubble detachment and coalescence behavior}  

Detachment radii align with Fritz's formula, demonstrating a linear relation between the volume-equivalent radii and the contact angles. 
Multi-bubble cases exhibit earlier detachment with reduced radii compared to isolated bubbles. 
It is worth noting that coalescence reduces the detachment time, thus potentially improving the mass transfer efficiency of the electrochemical system.

\noindent\textbf{Outlook}  

The present work focuses on single growth-detachment cycles due to computational constraints. 
Tracking deformable interfaces remains resource-intensive, limiting simulations to short time spans (\(t \sim0.2 \,\mathrm{s}\)). 
While comparable to prior studies using rigid bubble models \citep{sepahi2022a}, 
broader parametric investigations (e.g., current density variations, contact angle distributions) 
are necessary to strengthen statistical conclusions. For instance, a statistically steady Sherwood number at the electrode $\mathrm{Sh_e}$ has not been investigated in this work. Future efforts are required to develop a new scheme with a larger stable time step, which will help to enable long-term simulations. 

%Importantly, the experimental data available for comparison do not fully align with the objectives of this study. We will design and implement a more sophisticated experimental setup that will provide observational proof of our work (e.g., continuous single bubble detachment), thus increasing the reliability and impact of the three-dimensional numerical study.

\section*{Declaration of competing interest}
The authors declare that they have no known competing financial interests or personal relationships that could have appeared to influence the work reported in this paper.

\section*{Data availability}
Data will be made available on request.

\section*{Acknowledgements}
This project has received funding from the European Research Council (ERC) under the European Union’s Horizon 2020 research and innovation programme (grant agreement number 883849). 
We thank the French national GENCI supercomputing agency and the relevant supercomputer centers for their grants of CPU time on massively parallel machines, and their teams for assistance and the use of Irene-Rome at TGCC.
\FloatBarrier
\clearpage % <--- Force all floats and previous content to flush
\onecolumn % <--- Temporarily force single-column mode
\twocolumn % <--- Re-enter two-column mode, forcing the title to the top

% Now insert the bibliography content
\nobalance
\raggedbottom

\bibliographystyle{cas-model2-names}
\bibliography{library_1}

@article{lv2021,
  title = {{Self-Propelled Detachment upon Coalescence of Surface Bubbles}},
  author = {Lv, Pengyu and Pe{\~n}as, Pablo and Le The, Hai and Eijkel, Jan and {van den Berg}, Albert and Zhang, Xuehua and Lohse, Detlef},
  year = {2021},
  month = nov,
  journal = {Physical Review Letters},
  volume = {127},
  number = {23},
  pages = {235501},
  publisher = {American Physical Society},
  doi = {10.1103/PhysRevLett.127.235501},
  urldate = {2025-06-28},
  langid = {french}
}

@article{khalighi2023,
  title = {Hydrogen Bubble Growth in Alkaline Water Electrolysis: {{An}} Immersed Boundary Simulation Study},
  shorttitle = {Hydrogen Bubble Growth in Alkaline Water Electrolysis},
  author = {Khalighi, Faeze and Deen, Niels G. and Tang, Yali and Vreman, Albertus W.},
  year = {2023},
  month = mar,
  journal = {Chemical Engineering Science},
  volume = {267},
  pages = {118280},
  issn = {00092509},
  doi = {10.1016/j.ces.2022.118280},
  urldate = {2024-10-27},
  langid = {english}
}

@article{Holladay2009,
  title = {An Overview of Hydrogen Production Technologies},
  author = {Holladay, J. D. and Hu, J. and King, D. L. and Wang, Y.},
  year = {2009},
  month = jan,
  journal = {Catalysis Today},
  series = {Hydrogen {{Production}} - {{Selected}} Papers from the {{Hydrogen Production Symposium}} at the {{American Chemical Society}} 234th {{National Meeting}} \& {{Exposition}}, {{August}} 19-23, 2007, {{Boston}}, {{MA}}, {{USA}}},
  volume = {139},
  number = {4},
  pages = {244--260},
  issn = {0920-5861},
  doi = {10.1016/j.cattod.2008.08.039},
  urldate = {2024-05-22}
}

@article{Turner2004,
  title = {Sustainable {{Hydrogen Production}}},
  author = {Turner, John A.},
  year = {2004},
  month = aug,
  journal = {Science},
  volume = {305},
  number = {5686},
  pages = {972--974},
  publisher = {American Association for the Advancement of Science},
  doi = {10.1126/science.1103197},
  urldate = {2024-05-22}
}

@article{Dawood2020,
  title = {Hydrogen Production for Energy: {{An}} Overview},
  shorttitle = {Hydrogen Production for Energy},
  author = {Dawood, Furat and Anda, Martin and Shafiullah, G. M.},
  year = {2020},
  month = feb,
  journal = {International Journal of Hydrogen Energy},
  volume = {45},
  number = {7},
  pages = {3847--3869},
  issn = {0360-3199},
  doi = {10.1016/j.ijhydene.2019.12.059},
  urldate = {2024-05-22}
}

@article{Swiegers2021,
  title = {The Prospects of Developing a Highly Energy-Efficient Water Electrolyser by Eliminating or Mitigating Bubble Effects},
  author = {Swiegers, Gerhard F. and Terrett, Richard N. L. and Tsekouras, George and Tsuzuki, Takuya and Pace, Ronald J. and Stranger, Robert},
  year = {2021},
  month = mar,
  journal = {Sustainable Energy \& Fuels},
  volume = {5},
  number = {5},
  pages = {1280--1310},
  publisher = {The Royal Society of Chemistry},
  issn = {2398-4902},
  doi = {10.1039/D0SE01886D},
  urldate = {2024-05-22},
  langid = {english}
}

@article{Hreiz2015,
  title = {Electrogenerated Bubbles Induced Convection in Narrow Vertical Cells: {{A}} Review},
  shorttitle = {Electrogenerated Bubbles Induced Convection in Narrow Vertical Cells},
  author = {Hreiz, Rainier and Abdelouahed, Lokmane and F{\"u}nfschilling, Denis and Lapicque, Fran{\c c}ois},
  year = {2015},
  month = aug,
  journal = {Chemical Engineering Research and Design},
  volume = {100},
  pages = {268--281},
  issn = {02638762},
  doi = {10.1016/j.cherd.2015.05.035},
  urldate = {2024-05-30},
  langid = {english}
}

@article{Linde2018,
  title = {Gas Bubble Evolution on Microstructured Silicon Substrates},
  author = {Van der Linde, Peter and {Pe{\~n}as-L{\'o}pez}, Pablo and Soto, {\'A}lvaro Moreno and van der Meer, Devaraj and Lohse, Detlef and Gardeniers, Han and Rivas, David Fern{\'a}ndez},
  year = {2018},
  month = dec,
  journal = {Energy \& Environmental Science},
  volume = {11},
  number = {12},
  pages = {3452--3462},
  publisher = {The Royal Society of Chemistry},
  issn = {1754-5706},
  doi = {10.1039/C8EE02657B},
  urldate = {2024-05-22},
  langid = {english}
}

@article{Angulo2020,
  title = {Influence of {{Bubbles}} on the {{Energy Conversion Efficiency}} of {{Electrochemical Reactors}}},
  author = {Angulo, Andrea and {van der Linde}, Peter and Gardeniers, Han and Modestino, Miguel and Fern{\'a}ndez Rivas, David},
  year = {2020},
  month = mar,
  journal = {Joule},
  volume = {4},
  number = {3},
  pages = {555--579},
  issn = {2542-4351},
  doi = {10.1016/j.joule.2020.01.005},
  urldate = {2024-05-23}
}

@article{dapkus1986,
  title = {Nucleation of Electrolytically Evolved Hydrogen at an Ideally Smooth Electrode},
  author = {Dapkus, Kestutis V and Sides, Paul J},
  year = {1986},
  month = may,
  journal = {Journal of Colloid and Interface Science},
  volume = {111},
  number = {1},
  pages = {133--151},
  issn = {00219797},
  doi = {10.1016/0021-9797(86)90014-7},
  urldate = {2024-05-13},
  copyright = {https://www.elsevier.com/tdm/userlicense/1.0/},
  langid = {english}
}

@article{Jones1999,
  title = {The Cycle of Bubble Production from a Gas Cavity in a Supersaturated Solution},
  author = {Jones, S. F. and Evans, G. M. and Galvin, K. P.},
  year = {1999},
  month = feb,
  journal = {Advances in Colloid and Interface Science},
  volume = {80},
  number = {1},
  pages = {51--84},
  issn = {0001-8686},
  doi = {10.1016/S0001-8686(98)00075-X},
  urldate = {2024-05-23}
}

@article{Brandon1985,
  title = {Growth Kinetics of Bubbles Electrogenerated at Microelectrodes},
  author = {Brandon, N. P. and Kelsall, G. H.},
  year = {1985},
  month = jul,
  journal = {Journal of Applied Electrochemistry},
  volume = {15},
  number = {4},
  pages = {475--484},
  issn = {1572-8838},
  doi = {10.1007/BF01059288},
  urldate = {2024-05-23},
  langid = {english}
}

@article{Scriven1959,
  title = {On the Dynamics of Phase Growth},
  author = {Scriven, L. E.},
  year = {1959},
  month = apr,
  journal = {Chemical Engineering Science},
  volume = {10},
  number = {1},
  pages = {1--13},
  issn = {0009-2509},
  doi = {10.1016/0009-2509(59)80019-1},
  urldate = {2024-05-23}
}

@article{Epstein1950,
  title = {On the {{Stability}} of {{Gas Bubbles}} in {{Liquid}}-{{Gas Solutions}}},
  author = {Epstein, P. S. and Plesset, M. S.},
  year = {1950},
  month = nov,
  journal = {The Journal of Chemical Physics},
  volume = {18},
  number = {11},
  pages = {1505--1509},
  issn = {0021-9606},
  doi = {10.1063/1.1747520},
  urldate = {2024-05-23}
}

@article{Stephan1979,
  title = {A Model for Correlating Mass Transfer Data at Gas Evolving Electrodes},
  author = {Stephan, K. and Vogt, H.},
  year = {1979},
  month = jan,
  journal = {Electrochimica Acta},
  volume = {24},
  number = {1},
  pages = {11--18},
  issn = {0013-4686},
  doi = {10.1016/0013-4686(79)80033-X},
  urldate = {2024-05-23}
}

@article{El-Askary2015,
  title = {Hydrodynamics Characteristics of Hydrogen Evolution Process through Electrolysis: {{Numerical}} and Experimental Studies},
  shorttitle = {Hydrodynamics Characteristics of Hydrogen Evolution Process through Electrolysis},
  author = {{El-Askary}, W. A. and Sakr, I. M. and Ibrahim, K. A. and Balabel, A.},
  year = {2015},
  month = oct,
  journal = {Energy},
  volume = {90},
  pages = {722--737},
  issn = {0360-5442},
  doi = {10.1016/j.energy.2015.07.108},
  urldate = {2024-05-23}
}

@article{Soto2018,
  title = {Coalescence of Diffusively Growing Gas Bubbles},
  author = {Soto, {\'A}lvaro Moreno and Maddalena, Tom and Fraters, Arjan and van der Meer, Devaraj and Lohse, Detlef},
  year = {2018},
  month = jul,
  journal = {Journal of Fluid Mechanics},
  volume = {846},
  pages = {143--165},
  issn = {0022-1120, 1469-7645},
  doi = {10.1017/jfm.2018.277},
  urldate = {2024-05-23},
  langid = {english}
}

@article{Zhang2024,
  title = {Coalescence and Detachment of Double Bubbles on Electrode Surface in Photoelectrochemical Water Splitting},
  author = {Zhang, Bo and Wang, Yechun and Feng, Yuyang and Zhen, Canghao and Liu, Miaomiao and Cao, Zhenshan and Zhao, Qiuyang and Guo, Liejin},
  year = {2024},
  month = mar,
  journal = {Cell Reports Physical Science},
  volume = {5},
  number = {3},
  pages = {101837},
  issn = {26663864},
  doi = {10.1016/j.xcrp.2024.101837},
  urldate = {2024-05-23},
  langid = {english}
}

@article{Hawkes2009,
  title = {{{3D CFD}} Model of a Multi-Cell High-Temperature Electrolysis Stack},
  author = {Hawkes, Grant and O'Brien, James and Stoots, Carl and Hawkes, Brian},
  year = {2009},
  month = may,
  journal = {International Journal of Hydrogen Energy},
  volume = {34},
  number = {9},
  pages = {4189--4197},
  issn = {0360-3199},
  doi = {10.1016/j.ijhydene.2008.11.068},
  urldate = {2024-05-23}
}

@article{Abdelouahed2014,
  title = {Hydrodynamics of Gas Bubbles in the Gap of Lantern Blade Electrodes without Forced Flow of Electrolyte: {{Experiments}} and {{CFD}} Modelling},
  shorttitle = {Hydrodynamics of Gas Bubbles in the Gap of Lantern Blade Electrodes without Forced Flow of Electrolyte},
  author = {Abdelouahed, Lokmane and Hreiz, Rainier and Poncin, Souhila and Valentin, G{\'e}rard and Lapicque, Francois},
  year = {2014},
  month = may,
  journal = {Chemical Engineering Science},
  volume = {111},
  pages = {255--265},
  issn = {0009-2509},
  doi = {10.1016/j.ces.2014.01.028},
  urldate = {2024-05-23}
}

@article{Taqieddin2017,
  title = {Review-{{Physicochemical}} Hydrodynamics of Gas Bubbles in Two Phase Electrochemical Systems},
  author = {Taqieddin, Amir and Nazari, Roya and Rajic, Ljiljana and Alshawabkeh, Akram},
  year = {2017},
  journal = {Journal of the Electrochemical Society},
  volume = {164},
  number = {13},
  pages = {E448-E459},
  issn = {0013-4651},
  doi = {10.1149/2.1161713jes},
  langid = {english},
  pmcid = {PMC5935447},
  pmid = {29731515}
}

@article{Burdyny2017,
  title = {Nanomorphology-{{Enhanced Gas-Evolution Intensifies CO2 Reduction Electrochemistry}}},
  author = {Burdyny, Tom and Graham, Percival and Pang, Yuanjie and Dinh, Cao Thang and Liu, Min and Sargent, Edward and Sinton, David},
  year = {2017},
  month = mar,
  journal = {ACS Sustainable Chemistry \& Engineering},
  volume = {5},
  doi = {10.1021/acssuschemeng.7b00023}
}

@article{Bashkatov2024,
  title = {Performance {{Enhancement}} of {{Electrocatalytic Hydrogen Evolution}} through {{Coalescence-Induced Bubble Dynamics}}},
  author = {Bashkatov, Aleksandr and Park, Sunghak and Demirk{\i}r, {\c C}ayan and Wood, Jeffery A. and Koper, Marc T. M. and Lohse, Detlef and Krug, Dominik},
  year = {2024},
  month = apr,
  journal = {Journal of the American Chemical Society},
  volume = {146},
  number = {14},
  pages = {10177--10186},
  publisher = {American Chemical Society},
  issn = {0002-7863},
  doi = {10.1021/jacs.4c02018},
  urldate = {2024-05-23}
}

@article{Gennari2022,
  title = {A Phase-Change Model for Diffusion-Driven Mass Transfer Problems in Incompressible Two-Phase Flows},
  author = {Gennari, Gabriele and {Jefferson-Loveday}, Richard and Pickering, Stephen J.},
  year = {2022},
  month = sep,
  journal = {Chemical Engineering Science},
  volume = {259},
  pages = {117791},
  issn = {0009-2509},
  doi = {10.1016/j.ces.2022.117791},
  urldate = {2024-05-23}
}

@article{Zhang2023,
  title = {Minimum Current for Detachment of Electrolytic Bubbles},
  author = {Zhang, Yixin and Lohse, Detlef},
  year = {2023},
  month = nov,
  journal = {Journal of Fluid Mechanics},
  volume = {975},
  pages = {R3},
  issn = {0022-1120, 1469-7645},
  doi = {10.1017/jfm.2023.898},
  urldate = {2024-05-23},
  langid = {english}
}

@article{Glas1964,
  title = {Measurements of the Growth of Electrolytic Bubbles},
  author = {Glas, J. P. and Westwater, J. W.},
  year = {1964},
  month = dec,
  journal = {International Journal of Heat and Mass Transfer},
  volume = {7},
  number = {12},
  pages = {1427--1443},
  issn = {0017-9310},
  doi = {10.1016/0017-9310(64)90130-9},
  urldate = {2024-05-23}
}

@article{vogt2004,
  title = {The Limits of the Analogy between Boiling and Gas Evolution at Electrodes},
  author = {Vogt, H. and Aras, {\"O}. and Balzer, R.J.},
  year = {2004},
  month = feb,
  journal = {International Journal of Heat and Mass Transfer},
  volume = {47},
  number = {4},
  pages = {787--795},
  issn = {00179310},
  doi = {10.1016/j.ijheatmasstransfer.2003.07.023},
  urldate = {2024-07-10},
  copyright = {https://www.elsevier.com/tdm/userlicense/1.0/},
  langid = {english}
}

@article{fleckenstein2015,
  title = {A {{Volume-of-Fluid-based}} Numerical Method for Multi-Component Mass Transfer with Local Volume Changes},
  author = {Fleckenstein, Stefan and Bothe, Dieter},
  year = {2015},
  month = nov,
  journal = {Journal of Computational Physics},
  volume = {301},
  pages = {35--58},
  issn = {00219991},
  doi = {10.1016/j.jcp.2015.08.011},
  urldate = {2024-07-01},
  langid = {english}
}

@book{Tryggvason_Scardovelli_Zaleski_2011, place={Cambridge}, title={Direct Numerical Simulations of Gas–Liquid Multiphase Flows}, publisher={Cambridge University Press}, author={Tryggvason, Grétar and Scardovelli, Ruben and Zaleski, Stéphane}, year={2011}}

@article{vogt2012,
  title = {The Actual Current Density of Gas-Evolving Electrodes---{{Notes}} on the Bubble Coverage},
  author = {Vogt, H.},
  year = {2012},
  month = sep,
  journal = {Electrochimica Acta},
  volume = {78},
  pages = {183--187},
  issn = {00134686},
  doi = {10.1016/j.electacta.2012.05.124},
  urldate = {2024-07-10},
  copyright = {https://www.elsevier.com/tdm/userlicense/1.0/},
  langid = {english}
}

@article{vogt2005,
  title = {The Bubble Coverage of Gas-Evolving Electrodes in Stagnant Electrolytes},
  author = {Vogt, H. and Balzer, R.J.},
  year = {2005},
  month = mar,
  journal = {Electrochimica Acta},
  volume = {50},
  number = {10},
  pages = {2073--2079},
  issn = {00134686},
  doi = {10.1016/j.electacta.2004.09.025},
  urldate = {2024-07-10},
  copyright = {https://www.elsevier.com/tdm/userlicense/1.0/},
  langid = {english}
}

@article{fouad1972,
  title = {Effect of Gas Evolution on the Rate of Mass Transfer at Vertical Electrodes},
  author = {Fouad, M. G. and Sedahmed, G. H.},
  year = {1972},
  month = apr,
  journal = {Electrochimica Acta},
  volume = {17},
  number = {4},
  pages = {665--672},
  issn = {0013-4686},
  doi = {10.1016/0013-4686(72)80067-7},
  urldate = {2025-02-12}
}

@article{hine1980,
  title = {Bubble {{Effects}} on the {{Solution IR Drop}} in a {{Vertical Electrolyzer Under Free}} and {{Forced Convection}}},
  author = {Hine, Fumio and Murakami, Koichi},
  year = {1980},
  month = feb,
  journal = {Journal of The Electrochemical Society},
  volume = {127},
  number = {2},
  pages = {292},
  publisher = {IOP Publishing},
  issn = {1945-7111},
  doi = {10.1149/1.2129658},
  urldate = {2025-02-12},
  langid = {english}
}

@article{li2018,
  title = {In-Situ Investigation of Bubble Dynamics and Two-Phase Flow in Proton Exchange Membrane Electrolyzer Cells},
  author = {Li, Yifan and Kang, Zhenye and Mo, Jingke and Yang, Gaoqiang and Yu, Shule and Talley, Derrick A. and Han, Bo and Zhang, Feng-Yuan},
  year = {2018},
  month = jun,
  journal = {International Journal of Hydrogen Energy},
  volume = {43},
  number = {24},
  pages = {11223--11233},
  issn = {03603199},
  doi = {10.1016/j.ijhydene.2018.05.006},
  urldate = {2025-02-11},
  langid = {english}
}

@article{vanderlinde2017a,
  title = {Electrolysis-{{Driven}} and {{Pressure-Controlled Diffusive Growth}} of {{Successive Bubbles}} on {{Microstructured Surfaces}}},
  author = {{Van der Linde}, Peter and Moreno Soto, {\'A}lvaro and {Pe{\~n}as-L{\'o}pez}, Pablo and {Rodr{\'i}guez-Rodr{\'i}guez}, Javier and Lohse, Detlef and Gardeniers, Han and {van der Meer}, Devaraj and Fern{\'a}ndez Rivas, David},
  year = {2017},
  month = nov,
  journal = {Langmuir},
  volume = {33},
  number = {45},
  pages = {12873--12886},
  publisher = {American Chemical Society},
  issn = {0743-7463},
  doi = {10.1021/acs.langmuir.7b02978},
  urldate = {2024-05-29}
}

@article{sepahi2022a,
  title = {The Effect of Buoyancy Driven Convection on the Growth and Dissolution of Bubbles on Electrodes},
  author = {Sepahi, Farzan and Pande, Nakul and Chong, Kai Leong and Mul, Guido and Verzicco, Roberto and Lohse, Detlef and Mei, Bastian T. and Krug, Dominik},
  year = {2022},
  month = jan,
  journal = {Electrochimica Acta},
  volume = {403},
  pages = {139616},
  issn = {00134686},
  doi = {10.1016/j.electacta.2021.139616},
  urldate = {2024-03-27},
  langid = {english}
}

@article{francois2011,
  title = {Direct Measurement of Mass Transfer around a Single Bubble by Micro-{{PLIFI}}},
  author = {Francois, J. and Dietrich, N. and Guiraud, P. and Cockx, A.},
  year = {2011},
  month = jul,
  journal = {Chemical Engineering Science},
  volume = {66},
  number = {14},
  pages = {3328--3338},
  issn = {00092509},
  doi = {10.1016/j.ces.2011.01.049},
  urldate = {2025-02-14},
  copyright = {https://www.elsevier.com/tdm/userlicense/1.0/},
  langid = {english}
}

@article{dani2007,
  title = {Local Measurement of Oxygen Transfer around a Single Bubble by Planar Laser-Induced Fluorescence},
  author = {Dani, Adil and Guiraud, Pascal and Cockx, Arnaud},
  year = {2007},
  month = dec,
  journal = {Chemical Engineering Science},
  volume = {62},
  number = {24},
  pages = {7245--7252},
  issn = {00092509},
  doi = {10.1016/j.ces.2007.08.047},
  urldate = {2025-02-14},
  copyright = {https://www.elsevier.com/tdm/userlicense/1.0/},
  langid = {english}
}

@article{lafmejani2017,
  title = {{{VOF}} Modelling of Gas--Liquid Flow in {{PEM}} Water Electrolysis Cell Micro-Channels},
  author = {Lafmejani, Saeed Sadeghi and Olesen, Anders Christian and K{\ae}r, S{\o}ren Knudsen},
  year = {2017},
  month = jun,
  journal = {International Journal of Hydrogen Energy},
  volume = {42},
  number = {26},
  pages = {16333--16344},
  issn = {03603199},
  doi = {10.1016/j.ijhydene.2017.05.079},
  urldate = {2025-02-14},
  langid = {english}
}

@article{li2019,
  title = {In-Situ Investigation and Modeling of Electrochemical Reactions with Simultaneous Oxygen and Hydrogen Microbubble Evolutions in Water Electrolysis},
  author = {Li, Yifan and Yang, Gaoqiang and Yu, Shule and Kang, Zhenye and Mo, Jingke and Han, Bo and Talley, Derrick A. and Zhang, Feng-Yuan},
  year = {2019},
  month = oct,
  journal = {International Journal of Hydrogen Energy},
  volume = {44},
  number = {52},
  pages = {28283--28293},
  issn = {03603199},
  doi = {10.1016/j.ijhydene.2019.09.044},
  urldate = {2025-02-12},
  langid = {english}
}
\end{document}